\documentclass[aps,prb,twocolumn,showpacs,amsmath,amssymb,superscriptaddress,floatfix]{revtex4-1}
\usepackage{graphicx}
\usepackage{dcolumn}
\usepackage{bm}
\usepackage{amsfonts}
\usepackage{amsmath}
\usepackage{ulem}
\usepackage{color}
\usepackage{hyperref}
\begin{document}

\title{Tunnel junction of helical edge states: Determining and controlling spin-preserving and spin-flipping processes through transconductance} 

\author{Pietro Sternativo}
\affiliation{Dipartimento di Scienza Applicata e Tecnologia del Politecnico di Torino, I-10129 Torino, Italy}
\author{Fabrizio Dolcini}
\email{fabrizio.dolcini@polito.it}

\affiliation{Dipartimento di Scienza Applicata e Tecnologia del Politecnico di Torino, I-10129 Torino, Italy}

\affiliation{CNR-SPIN, Monte S.Angelo - via Cinthia, I-80126 Napoli, Italy}

\begin{abstract}
When a constriction is realized in a 2D quantum spin Hall system, electron tunneling between  helical edge states occurs via two types of channels allowed by time-reversal symmetry, namely spin-preserving ({p}) and spin-flipping ({f}) tunneling processes. Determining and controlling the effects of these two channels is crucial to the application of helical edge states in spintronics. We show that, despite the Hamiltonian terms describing these two processes do not commute, the scattering matrix entries of the related 4-terminal setup  always factorize into products of p-terms  and  f-terms contributions. Such factorization provides an operative way to determine the transmission coefficient $T_p$ and $T_f$ related to each of the two processes, via transconductance measurements. 
Furthermore, these transmission coefficients are also found to be controlled independently by a suitable combination of  two gate voltages applied across the junction. This result holds for an arbitrary profile of the tunneling amplitudes, including   disorder in the tunnel region, enabling us to discuss the effect of the finite length of the tunnel junction, and the space modulation of both magnitude and phase of the tunneling amplitudes.
\end{abstract}


\maketitle

%%%%%%%%%%%%%%%%%%%%%%%%%%%%%%%%%%%%%%%%%%%%%%%%%%%%%%%%%%%%%%%%%%%%%%%%%%%%%%%%%%%%
%%%%%%%%%%%%%%%%%%%%%%%%%%%%%%%%%%%%%%%%%%%%%%%%%%%%%%%%%%%%%%%%%%%%%%%%%%%%%%%%%%%%
%%%%%%%%%%%%%%%%%%%%%%%%%%%%%%%%%%%%%%%%%%%%%%%%%%%%%%%%%%%%%%%%%%%%%%%%%%%%%%%%%%%%
\section{Introduction}
The discovery of topological materials~\cite{TM-1-theo,TM-1-exp,helical-image,TM-2,TM-reviews} has unveiled the existence of helical edge states, i.e. linearly dispersed gapless one-dimensional modes flowing at the boundaries of the insulating bulk of a two-dimensional Quantum Spin Hall effect system~\cite{TM-1-theo,TM-1-exp,helical-image,zhang-PRL,InAsGaSb}. Helical states are characterized by a locking of the electron group velocity to the spin orientation, so that the two counter-propagating modes flowing at a given edge exhibit opposite spin orientation.   
The most straightforward way to reveal the peculiar features of helical edge states is through their transport properties, in particular when the helical states of opposite edges are coupled in a tunnel junction, realized e.g. by etching a  constriction in   HgTe/CdTe~\cite{TM-1-exp,helical-image} or in InAs/GaSb~\cite{InAsGaSb} quantum wells. 
In such situation, time-reversal symmetry implies that two  types  of  tunnel couplings between helical edge states exist~\cite{zhang-PRL,teo}. The first type~({p}) preserves the electron spin and changes the group velocity, similarly to a  backscattering term also present in conventional quantum wires. The second type~({f}) instead induces spin flipping by preserving the group velocity. One may expect that the coupling constant of the spin-flipping processes is smaller than that of the spin-preserving ones. However, no operative way has been conceived so far to quantitatively extract these coupling constants. In fact, the possibility of exploiting both processes is   crucial  for the intriguing perspective of utilizing topological edge states in spintronics\cite{spintronics,spintronics-TI,richter,sassetti-cavaliere-2013}, where the spin degree of freedom is used to encode information. Indeed in a spintronics nanodevice  currents should be switched on demand from one terminal to another, in a controlled way, with either preservation or flipping of the spin, depending on the specific operation to be performed. 

\begin{figure} 
\centering
\includegraphics[width=8cm,clip]{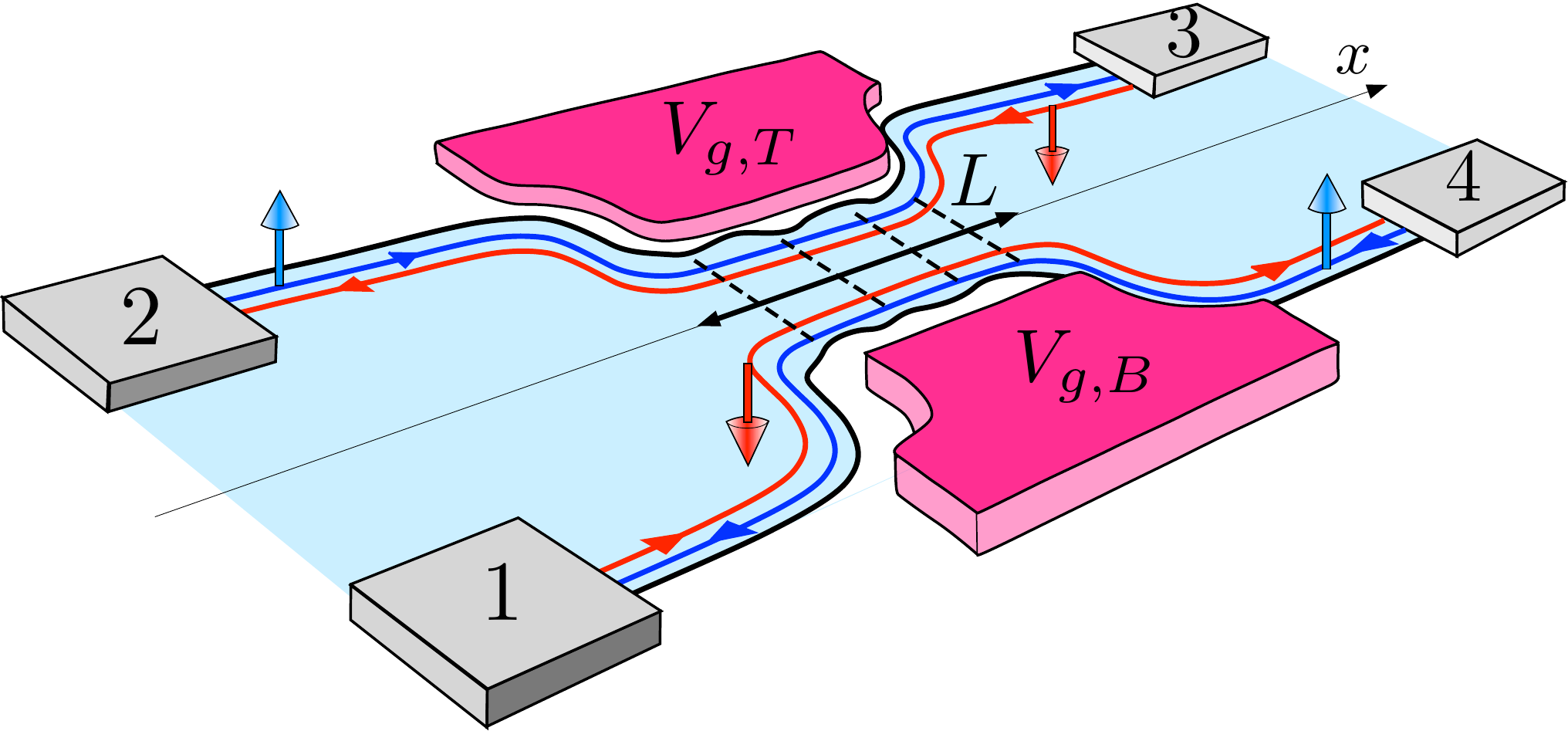}
\caption{(Color online) A tunnel junction coupling the helical edge states can be realized by etching a constriction in a four terminal Quantum Spin Hall effect setup. Two side gates enable one to tune the potential along the junction. Two types of time-reversal symmetric tunnel processes occur at the junction, namely spin-preserving   and  spin-flipping processes. Determining and controlling these two processes is crucial for exploiting helical edge states in spintronics devices.}
\label{Fig-01}
\end{figure}

Determining and controlling the transmission coefficients related to these two tunneling channels  appears to be a challenging   task, for various reasons.   In the first instance  the Hamiltonian terms describing the  two  processes do not commute, making the analysis of their interplay a priori non-trivial. 
Secondly, due to etching and to indirect coupling via the bulk states, a tunnel junction is a typically irregular and disordered region, implying that the tunneling amplitude of each of the two channels cannot be described by one single parameter but rather by a space-dependent profile. Furthermore, in a tunnel junction realized in a quantum well of materials like HgTe/CdTe, where  the strong spin-orbit interaction  correlates momentum and spin~\cite{TM-1-exp,SO-HgTe}, the breaking of the longitudinal transversal invariance originating from the finite length of the tunnel region affects spin preservation, thereby possibly modifying the relative weight of {f}-processes   with respect to {p}-processes.

Most of the theoretical approaches to this problem have treated the tunnel junction as a point-like constriction, using a $\delta$-tunneling (DT) model, and have focussed on the effect of electron-electron interaction on conductance~\cite{teo,chamon_2009,strom_2009,trauz-recher,dolcini2012,dolcetto-sassetti} and noise~\cite{noise}, and on   interference phenomena between two quantum point contacts~\cite{akhmerov,dolcini2011,virtanen-recher,citro-romeo,citro-sassetti,liliana}. However, while there is no clear experimental evidence that electron-electron interaction plays a significant role in helical edge states transport, it is worth noticing that the DT model, per se, completely neglects the internal structure of the junction, which may contain important physical insights in  realistic implementations. Some recent works, for instance, have pointed out that the finite length of the junction  plays an important role when the related Thouless energy becomes comparable to the applied voltages~\cite{trauz-recher,richter,citro-sassetti,dolcetto-sassetti}. These studies were limited to specific profiles  for the   tunneling amplitudes and/or to the case of spin-preserving tunneling, though. \\

For these reasons, the question how to operatively determine and   control the spin-preserving and spin-flipping effects in a realistic   tunnel junction in the presence of a disordered   profile is still open. 
In this paper we address this problem. Focussing on the regime where electron-electron interaction is negligible, we consider a 4-terminal setup, schematically depicted in Fig.\ref{Fig-01}, where helical edge states are coupled in a  tunnel  region characterized by  a finite length and by an arbitrary profile for the tunneling amplitude of both {p}- and {f}-tunneling processes. We show that, despite the Hamiltonian terms describing the two types of tunneling processes do not commute, the scattering matrix  entries of the 4-terminal setup always factorize into two terms that depend on spin-preserving terms only and on spin-flipping terms only. This factorization   provides an operative way to determine the magnitude of the transmission coefficients related to each of these terms, via transport measurements. In particular, the spin-flipping terms, although possibly quantitatively smaller in magnitude than the spin-preserving terms, induce qualitatively different features in  the conductance matrix, which cannot be ascribed to {p}-processes.    Furthermore we predict that, by suitably operating with side gates, one can control independently the two transmission coefficients related to spin-preserving and spin-flipping tunneling.  
Importantly, the factorization property holds for an {\it arbitrary} profile of the tunneling amplitudes, possibly including disorder and local fluctuations effects. In particular, it also holds in the limit of short constriction. By comparing our results with the widely used DT model,  we shall show that such factorization property, which is seemingly absent in DT model, is simply hidden in a misleading parametrization of the coupling constants of that model.\\

The paper is organized as follows: in Sec.\ref{sec-II}, after presenting the model for the tunnel junction, we prove   the factorization property of the scattering matrix  for an arbitrary profile for the junction parameters. In Sec.\ref{sec-III} we discuss how such property impacts the multi-terminal conductance, and show how the transmission coefficients related to the two types of tunneling can be operatively determined and controlled by gate voltages. In Sec.\ref{sec-IV} we present some explicit results for specific profiles of the junction tunneling amplitude, which enable us to discuss the role of {p}- and {f}-tunneling, the effects of the finite length of the tunnel junction and the variation of the magnitude and the phase of the tunneling amplitude profile. We also explain why the DT model hides the factorization property. Finally, in Sec.\ref{sec-V} we discuss our results and draw some conclusions.

%%%%%%%%%%%%%%%%%%%%%%%%%%%%%%%%%%%%%%%%%%%%%%%%%%%%%%
%%%%%%%%%%%%%%%%%%%%%%%%%%%%%%%%%%%%%%%%%%%%%%%%%%%%%%
%%%%%%%%%%      S e c t i o n   II    %%%%%%%%%%%%%%%%
%%%%%%%%%%%%%%%%%%%%%%%%%%%%%%%%%%%%%%%%%%%%%%%%%%%%%%
%%%%%%%%%%%%%%%%%%%%%%%%%%%%%%%%%%%%%%%%%%%%%%%%%%%%%%
 
\section{Model}
\label{sec-II}
We consider a 4-terminal setup of helical edge states, as sketched in Fig.\ref{Fig-01}. Along the Top edge of the quantum well right movers are characterized by spin $\uparrow$  and left  movers by spin $\downarrow$, whereas along the Bottom edge  the opposite spin orientations occur. A constriction is assumed to be realized in the quantum well,  allowing electron tunneling between the four edge states  over a region extending along the longitudinal direction $x$. Furthermore, we consider two gate electrodes, applied at the  sides of the constriction, enabling to shift the chemical potential of the edge states.
\\

We describe the  electron edge states through  four  electron field operators
$\Psi_{R\uparrow}(x) ,
\Psi_{L\uparrow}(x) ,
\Psi_{R\downarrow}(x) ,
\Psi_{L\downarrow}(x)
$,  with $\alpha=R/L=\pm$ denoting the chirality for right- and left- movers, respectively, and $\sigma=\uparrow,\downarrow$  the spin component. We then model the setup by the following low-energy Hamiltonian  
\begin{equation}
\hat{\mathcal{H}} =\hat{\mathcal{H}}_{0}   \, + \,   
  \hat{\mathcal{H}}_{g,T}\, + \,   
  \hat{\mathcal{H}}_{{g,B}}  \, + \,       
  \hat{\mathcal{H}}_{\rm tun}^{p} \, + \,    \hat{\mathcal{H}}_{\rm tun}^{f} \quad,   \label{HAM-FER}
\end{equation}
where  
\begin{equation}
\hat{\mathcal{H}}_0 = -i    \hbar v_{\rm F} \hspace{-0.3cm} \sum_{\alpha=R/L=\pm} \hspace{-0.3cm}\alpha \sum_{\sigma=\uparrow, \downarrow}\int   \! dx \,   :   \Psi^{\dagger}_{\alpha \sigma}(x)\, \partial_x \Psi^{}_{\alpha \sigma}(x)    :                 
\label{H0}
\end{equation}
%\begin{eqnarray}
%\hat{\mathcal{H}}_0 &=& -i    \hbar v_{\rm F} \sum_{\sigma=\uparrow, \downarrow}\!  \int   \! dx \,  \left[:   \Psi^{\dagger}_{R \sigma}(x)\, \partial_x \Psi^{}_{R \sigma}(x):  -  \right. \nonumber \\
%& & \hspace{2.5cm} \left. -:\Psi^{\dagger}_{L\bar{\sigma}}(x) \, \partial_x \Psi^{}_{L \bar{\sigma}}(x) \, :\right] \,                
%\label{H0}
%\end{eqnarray}
describes the linear bands of the helical edge states~\cite{TM-1-theo,TM-1-exp} ($: \, \, \, :$  denotes normal ordering), and
\begin{eqnarray}
\hat{\mathcal{H}}_{g,T} &=& \! \!    \displaystyle  \! \int    \! dx     \, eV_{g,T}(x) \left(   \hat{\rho}_{R \uparrow}(x)\,     \,+  \hat{\rho}_{L \downarrow}(x)\,  \right) \label{HgateT}  \\
\hat{\mathcal{H}}_{g,B} &=&  \displaystyle  \! \int    \! dx \, eV_{g,B}(x) \left(   \hat{\rho}_{R \downarrow}(x)    \,+ \,  \hat{\rho}_{L \uparrow}(x)   \right) 
\label{HgateB} 
\end{eqnarray}
account for the the electric potentials applied by the side gates across the constriction. Here 
\begin{equation} \label{rho-def}
\hat{\rho}_{\alpha \sigma}(x)=\, :\Psi^\dagger_{\alpha \sigma}(x) \Psi^{}_{\alpha \sigma}(x):
\end{equation}
is the electron chiral density. 
%%%%%%%%%%%%%%%%%%%%%%%%%%%%%%%%%%%%%%%%%%%%%%%%%
Finally, the last two terms in Eq.(\ref{HAM-FER}),   
\begin{equation}
\hat{\mathcal{H}}_{\rm tun}^{p} =\! \!    \displaystyle \sum_{\sigma=\uparrow, \downarrow}\! \int   \! dx      \left( \Gamma_{p}^{}(x) \,  \Psi^{\dagger}_{L \sigma}(x)\, \Psi^{}_{R \sigma}(x)    +   \mbox{h.c.} \right) \, \;  
\label{Hsp} 
\end{equation}
%%%%%%%%%%%%%%%%%%%%%%%%%%%%%%%%%%%%%%%%%%%%%%%%%
\begin{equation}
\hat{\mathcal{H}}_{\rm tun}^{f} =\hspace{-0.4cm} \sum_{\alpha=R/L=\pm} \hspace{-0.3cm} \alpha \int   \! dx    \,  \left(   \Gamma_{f}^{}(x) \, \,  \Psi^{\dagger}_{\alpha \downarrow}(x)\, \Psi^{}_{\alpha \uparrow}(x)   \, +     \mbox{h.c.} \right)    \label{Hsf} 
\end{equation}
describe  the two types of inter-edge tunnel coupling  allowed by time-reversal symmetry at the constriction, namely the spin-preserving and the  spin-flipping tunneling, respectively~\cite{zhang-PRL,teo,richter,trauz-recher,citro-romeo,dolcini2011,citro-sassetti,dellanna}. \\
%%%%%%%%%%%%%%%%%%%%%%%%%%%%%%

The tunneling amplitudes $\Gamma_{p}(x) \, , \Gamma_{f}(x)$ and the side gate potentials $V_{g,T}(x),V_{g,B}(x)$ characterizing the constriction region are allowed to vary along the longitudinal direction $x$. Their profiles can  be arbitrary, with the only constraint that --sufficiently  far away from the central region-- they   all vanish  and the helical states are described by the linear dispersion band term~(\ref{H0}) only. We can thus assume, without loss of generality, that there exist two coordinates $x_1$ and~$x_2$ defining the extremal longitudinal boundaries of the constriction, such that
\begin{equation}
\Gamma_{p}(x), \Gamma_{f}(x), V_{g,T}(x)\, , V_{g,B}(x) \, =0 \hspace{0.4cm} \mbox{for}\, \left\{ \begin{array}{l} x \le x_1  \\   \\ x\ge x_2\end{array} \,  \right., \label{vanish}
\end{equation}
with $L=x_2-x_1$ denoting the tunnel junction length.  
The tunnel junction $x_1 \le x \le x_2$ can thus be regarded to as the scattering region  in  the 4-terminal setup in Fig.\ref{Fig-01}, where the distributions of the incoming electrons are controlled by the chemical potentials $\mu_i$ ($i=1,\ldots 4$), and by the temperature $k_B T$ of the four reservoirs (see also Fig.\ref{Fig-02-gedankenexp}).\\

It is worth emphasizing that the two types of tunneling terms~(\ref{Hsp}) and (\ref{Hsf}) acting in the constriction do {\it not} commute,
\begin{eqnarray}
\left[ \hat{\mathcal{H}}_{\rm tun}^{p} \, , \, \hat{\mathcal{H}}_{\rm tun}^{f} \right] =  
  2 \int \! dx \, \left[   \Gamma^{}_{p}(x) \Gamma^{}_{f}(x) \Psi^\dagger_{L \downarrow}(x) \Psi^{}_{R \uparrow}(x)\, +\nonumber \right.  \\
 \left. +\Gamma^{}_{p}(x) \Gamma^{*}_{f}(x) \Psi^\dagger_{L \uparrow}(x) \Psi^{}_{R \downarrow}(x)  \,  - \, \mbox{h.c.}  \right] \neq 0 \hspace{1cm} \label{non-comm}
\end{eqnarray}
In view of Eq.(\ref{non-comm}), the terms $\hat{\mathcal{H}}_{\rm tun}^{p}$ and $\hat{\mathcal{H}}_{\rm tun}^{f}$ are expected to interplay in transport measurements, so that singling out the effect  of each of the two tunnel couplings looks quite difficult. This expectation seems to be  confirmed by transport predictions based on the DT model, which treats the tunnel junction as a point-like tunneling region~\cite{teo,chamon_2009,strom_2009,trauz-recher,dolcini2012,dolcetto-sassetti,noise,akhmerov,dolcini2011,virtanen-recher,citro-romeo,citro-sassetti,liliana}. Indeed by adopting such model the transmission coefficients are found depend on both tunneling amplitudes in a non-factorized way~\cite{dolcini2011,citro-romeo}.\\
 
We shall show that such seemingly complicated dependence is an artifact of the DT model. Indeed  we  prove that, despite the non-commutativity $\hat{\mathcal{H}}_{\rm tun}^{p}$ and $\hat{\mathcal{H}}_{\rm tun}^{f}$,   each entry of the scattering matrix of the setup can be factorized, one related the spin-preserving   and the other one to the spin-flipping tunneling. 
A comparison of our results with the ones of the DT model will be explicitly made in Sec.\ref{sec-IV-DT}.\\

%%%%%%%%%%%%%%%%%%%%%%%%%%%%%%%%%%%%%%%%%%%%%%%%%%%%%
\subsection{Factorization of the Scattering Matrix entries}
\label{sec-II-a}

In order to prove the factorization of the Scattering matrix entries, we first introduce  a four component electron field operator
$
\Psi(x) =\left( 
\Psi_{R\uparrow}(x) ,
\Psi_{L\uparrow}(x) ,
\Psi_{R\downarrow}(x) ,
\Psi_{L\downarrow}(x) 
\right)^T
$, as well as  Pauli matrices $\boldsymbol\sigma=(\sigma_x,\sigma_y, \sigma_z)$ acting on the spin space ($\sigma=\uparrow,\downarrow$),  and Pauli matrices $\boldsymbol\tau=(\tau_x,\tau_y, \tau_z)$ acting on the chirality space ($\alpha=R,L$). Furthermore we
define charge and  spin gate voltages as sum and difference of the side gate voltages appearing in Eqs.(\ref{HgateT})-(\ref{HgateB})
\begin{equation}
V_{g \,c/s}(x) =  (V_{g,T}(x)\pm V_{g,B}(x)) \, / 2  \quad. \label{Vgcs-def}
\end{equation}
The Hamiltonian (\ref{HAM-FER}) is then compactly written as
\begin{eqnarray}
\label{H-m}
\hat{\mathcal{H}}&=&\int \, dx \, \Psi^{\dagger}(x) \, \left[H_0(x) +H_{gc}(x)+H_{gs}(x)+ \right. \nonumber \\
& & \hspace{2cm} \left.  + H^{p}_{\rm tun}(x)+H^{f}_{\rm tun}(x)\right] \Psi^{}(x)
\end{eqnarray}
where the $4\times 4$ matrices in Eq.(\ref{H-m}) read 
\begin{eqnarray}\label{Hterms_def}
H_0(x) \, &=& -i\hbar v_F \left(\sigma_0 \otimes \tau_z \right)   \partial_x  \label{H0-m} \\
H_{gc}(x) &=& \displaystyle eV_{gc}(x)  \left( \sigma_0 \otimes \tau_0 \right)  \label{Hgc-m} \\
H_{gs}(x) &=& \displaystyle  eV_{gs}(x)  \left(  \sigma_z \otimes \tau_z \right)  \label{Hgs-m}\\
H^{p}_{\rm tun}(x) &=& \displaystyle    |\Gamma_{p}(x)|   \, \sigma_0 \otimes \left( \tau_x \cos\phi_{p}(x) + \tau_y \sin\phi_{p}(x) \right)    \, \,  \,   \label{Hp-m} 
\\
H^{f}_{\rm tun}(x) &=&    |\Gamma_{f}(x)|   \left( \sigma_x \cos\phi_{f}(x) + \sigma_y \sin\phi_{f}(x) \right)  \otimes \tau_z    \hspace{0.5cm} \label{Hf-m}  
\end{eqnarray}
with $\sigma_0$ and $\tau_0$ denoting  the $2 \times 2$ identity matrices in spin and chirality space, respectively. \\

The equation of motion $i\hbar \, \partial_t  \Psi(x,t)=[ \Psi(x,t) \,, \hat{\mathcal{H}}]$ obtained from the Hamiltonian (\ref{H-m}) implies for the stationary solutions $\Psi(x,t)=e^{-i E t/\hbar} \Psi_E(x)$ that
\begin{eqnarray}
 \lefteqn{i(\sigma_0  \otimes \tau_0)\,\frac{\partial }{\partial x}   \Psi_E(x)  
= } & &\label{eom-psi} \\
& & =  \left(   ( \boldsymbol\sigma \cdot \mathbf{b}_f(x) )\otimes \tau_0   \, +\sigma_0 \otimes ( \boldsymbol\tau \cdot \mathbf{b}_{p,E}(x)  )  \,  \right)  \Psi_E(x)  \nonumber
\end{eqnarray}
where 
\begin{eqnarray}
\mathbf{b}_f(x)= \left({\rm Re}  \Gamma_{f}(x)  ,  {\rm Im} \Gamma_{f}(x) ,  eV_{gs}(x) \right) / \hbar v_F \hspace{1cm} \label{bf-def}
\end{eqnarray}
is a (real) vector field that depends on the spin-flipping tunneling amplitude $\Gamma_{f}$ and the spin gate voltage $V_{gs}$, whereas
\begin{eqnarray}
\lefteqn{\mathbf{b}_{p,E}(x) =}  & & \label{bE-def} \\
&=& (-i\, {\rm Im}  \Gamma_{p}(x)   \, , \,  i\, {\rm Re}  \Gamma_{p}(x)   \, , \,\,  eV_{gc}(x) -E) / \hbar v_F \nonumber 
\end{eqnarray}
is a (complex) vector field that depends on the spin-preserving tunneling amplitude $\Gamma_{p}$, the charge gate $V_{gc}$ and the energy $E$, defined with respect to the Dirac point level.  Notice that Eq.~(\ref{eom-psi}) is formally equivalent to the `evolution' equation of a particle endowed with a twofold spin degree of freedom,   exposed to two `time-dependent' magnetic fields $\mathbf{b}_{f}$ and $\mathbf{b}_{p,E}$. The   role of time is played by space $x$, and the two magnetic fields originate from spin-preserving and spin-flipping tunneling.  
Outside the central scattering region, where Eq.(\ref{vanish}) holds, the  dynamics is governed by the term (\ref{H0-m}) only, and the field operator $\Psi_E$ solving Eq.(\ref{eom-psi})   acquires the simple asymptotic form 
\begin{equation}
\label{psi-asympt-x1}
\Psi_E(x \le x_1)=   (\sigma_0 \otimes e^{i \tau_z k_Ex} ) \,\left(   {a}_{R\uparrow}, {b}_{L\uparrow} , {a}_{R\downarrow},  {b}_{L\downarrow}  \right)^T
\end{equation}
and 
\begin{equation}
\label{psi-asympt-x2}
\Psi_E(x \ge x_2)=   (\sigma_0 \otimes e^{i \tau_z k_Ex} ) \,\left(   {b}_{R\uparrow}, {a}_{L\uparrow} , {b}_{R\downarrow},  {a}_{L\downarrow} \right)^T
\end{equation}
where $a_{\alpha \sigma}$ and $b_{\alpha \sigma}$ denote operators for incoming and outgoing states, respectively, and $k_E=E/\hbar v_F$.   The transfer Matrix ${\mathsf{M}}$,  connecting operators on the right of the central scattering region to the ones on the left,   
\begin{equation}
\label{M-def}
\left( \begin{array}{c}  {b}_{R\uparrow}\\  {a}_{L\uparrow} \\  {b}_{R\downarrow}\\  {a}_{L\downarrow} \end{array} \right) \, = \, {\mathsf{M}} \, \left( \begin{array}{c}  {a}_{R\uparrow}\\  {b}_{L\uparrow} \\  {a}_{R\downarrow}\\  {b}_{L\downarrow} \end{array} \right)
\end{equation}
can therefore be evaluated through the relation 
\begin{equation}
\label{psix2-M-psix1}
\Psi_E(x_2)= (\sigma_0 \otimes e^{i \tau_z k_E x_2} ) \, {\mathsf{M}} \, (\sigma_0 \otimes e^{-i \tau_z k_E  x_1} ) \, \Psi_E(x_1) \, \, .
\end{equation}
In order to determine the solution of the stationary Eq.(\ref{eom-psi}), we introduce the following space `evolution' operators 
\begin{eqnarray}
\mathbf{U}_f(x;0)&= &   {\rm T} \,e^{\displaystyle -i \int_{0}^x \, dx^\prime  \boldsymbol\sigma   \cdot \mathbf{b}_f(x^\prime)   } \label{Uf-expr}\\
\mathbf{U}_{p,E}(x;0)&= &  {\rm T} \,e^{\displaystyle -i \int_{0}^x \, dx^\prime  \boldsymbol\tau  \cdot \mathbf{b}_{p,E}(x^\prime) \,    } \label{Up-expr}
\end{eqnarray}
that appear as two continuous sets of rotations, around the local `magnetic' field $\mathbf{b}_f$ (spin space) and `pseudo-magnetic'  field $\mathbf{b}_{p,E}$ (chirality space) determined by the tunnel junction parameter profiles [see Eqs.(\ref{bf-def}) and (\ref{bE-def}), respectively].  Here ${\rm T}$ denotes the `time' (actually space) ordering operator, and the `evolution' is with respect to the space origin $x=0$. Then Eqs.(\ref{Uf-expr})-(\ref{Up-expr}) lead to
\begin{eqnarray}
i \, \partial_x \mathbf{U}_f(x;0) &= & (\boldsymbol\sigma   \cdot \mathbf{b}_f(x) )  \,  \mathbf{U}_f(x;0)\label{evo-1} \\
i \, \partial_x \mathbf{U}_{p,E}(x;0) &= &( \boldsymbol\tau   \cdot \mathbf{b}_{p,E}(x)   )\,  \mathbf{U}_{p,E}(x;0)\label{evo-2} \quad.
\end{eqnarray}
Using Eqs.(\ref{evo-1}) and (\ref{evo-2}),   the solution of Eq.(\ref{eom-psi}) is straightforwardly verified to be
\begin{equation}
\label{sol}
\Psi_E(x)=\left(\mathbf{U}_f(x;0) \otimes \mathbf{U}_{p,E}(x;0) \right) \,  \Psi(0)\quad,  
\end{equation}
where $\Psi(0)$ is a four-component field operator at the space origin.
The   solution (\ref{sol}) then implies that
\begin{eqnarray}
\Psi_E(x_2) &=& \left(\mathbf{U}_f(x_2;x_1) \otimes \mathbf{U}_{p,E}(x_2;x_1) \right)  \Psi_E(x_1)  \label{ultima}
\end{eqnarray} 
Comparing Eqs.(\ref{psix2-M-psix1}) and (\ref{ultima}) one obtains
\begin{equation}
{\mathsf{M}}=  \mathbf{m}_f  \,   \otimes \, \mathbf{m}_p    \label{M-factorized} 
\end{equation} 
where
\begin{eqnarray}
\mathbf{m}_f & \doteq & {\rm T} \,e^{-i \int_{x_1}^{x_2} \, dx^\prime  \boldsymbol\sigma   \cdot \mathbf{b}_f(x^\prime)   }  \label{mf-def}\\
\mathbf{m}_p & \doteq &  e^{-i \tau_z k_E x_2}  \left( {\rm T} \,e^{-i \int_{x_1}^{x_2} \, dx^\prime  \boldsymbol\tau   \cdot \mathbf{b}_{p,E}(x^\prime)   }  \right)  e^{+i \tau_z k_E x_1} \, \, \, \label{mp-def}
\end{eqnarray}
We notice that, because $\mathbf{b}_f(x) \in \mathbb{R}$, Eq.(\ref{mf-def})  implies that $\mathbf{m}_f\in {\rm SU}(2)$. In contrast, because $\mathbf{b}_{p,E}(x) \in \mathbb{C}$,  $\mathbf{m}_p\notin {\rm SU}(2)$. However,  ${\rm det} \, \mathbf{m}_p=+1$, since $\boldsymbol\tau   \cdot \mathbf{b}_{p,E}$ is traceless. Equation~(\ref{M-factorized}) shows that the Transfer Matrix  is the tensor product of a $2\times 2$ matrix $\mathbf{m}_f$ acting on spin space by a $2 \times 2$ matrix $\mathbf{m}_p$ acting on chirality space. This property reflects on the structure of the Scattering Matrix ${\mathsf{S}}$, which expresses outgoing operators in terms of incoming operators~\cite{multi},
\begin{equation}
\left( \begin{array}{c}  {b}_{L\uparrow}\\  {b}_{L\downarrow} \\  {b}_{R\uparrow}\\  {b}_{R\downarrow} \end{array} \right) \, = \, {\mathsf{S}} \, \left( \begin{array}{c}  {a}_{R\downarrow}\\  {a}_{R\uparrow} \\  {a}_{L\downarrow}\\  {a}_{L\uparrow} \end{array} \right) \quad,
\end{equation}
and which can straightforwardly be obtained from the relations (\ref{M-def}). 
Exploiting Eq.(\ref{M-factorized}) and the properties ${\rm det}\, \mathbf{m}_p={\rm det}\,\mathbf{m}_f=+1$, one obtains
\begin{equation}
{\mathsf{S}}= \left( \begin{array}{cccc}
0 & r_{p}  &  t_{p} \, r^*_{f}  & t_{p} \, t^*_{f}  \\ & & & \\
%%%
 r_{p} & 0 & t_{p} \,t_{f} & -t_{p} \, r_{f} \\ & & & \\
%%%
- t_{p} \, r^*_{f} & t_{p} \, t_{f} & 0 & r^\prime_{p}  \\ & & & \\
%%%
t_{p} \, t^*_{f} & t_{p} \, r_{f} & r^\prime_{p} & 0
\end{array}\right) \quad.\label{S-res}
\end{equation} \\
In Eq.(\ref{S-res}) the quantities    
\begin{equation}
\left\{ \begin{array}{lcl}
t_{p} &=&  \displaystyle \frac{1}{(\mathbf{m}_p)_{22}}  \\ & & \\
r_{p} &=& \displaystyle - \frac{(\mathbf{m}_p)_{21}}{(\mathbf{m}_p)_{22}} =- \frac{(\mathbf{m}_p)_{12}^*}{(\mathbf{m}_p)_{22}} \quad, \end{array} \right.  \label{rel-mp}
\end{equation} 
and $r^\prime_{p}=-r^*_{p} \,  t_{p}/t^*_{p}$ are  determined by $\mathbf{m}_p$. They depend on the spin-preserving tunneling amplitude $\Gamma_{p}$ and on the charge gate voltage $V_{gc}$ only, besides the energy $E$ [see Eqs.(\ref{mp-def}) and (\ref{bE-def})].  
In contrast, in Eq.(\ref{S-res}) the quantities
\begin{equation}
\left\{ \begin{array}{lcl}
t_{f} &=&  \displaystyle  (\mathbf{m}_f)_{11} = (\mathbf{m}_f)_{22}^*  \\ & & \\
r_{f} &=& \displaystyle  (\mathbf{m}_{f})_{21}=-(\mathbf{m}_{f})_{12}^*    \end{array} \right. \label{rel-mf}
\end{equation} 
are determined by $\mathbf{m}_f$, and depend   on the spin-flipping tunneling amplitude $\Gamma_{f}$ and on the spin gate voltage $V_{gs}$ only [see Eqs.(\ref{mf-def}) and (\ref{bf-def})].   \\
 
Equation (\ref{S-res}) shows that, in a tunnel junction of helical edge states, the entries of the scattering matrix always   factorize  into products of two reflection and/or transmission amplitudes, one related to {p}-tunneling processes and the other one to {f}-tunneling processes. Such result holds for an {\it arbitrary} profile of the tunneling amplitudes $\Gamma_{p}(x)$ and $\Gamma_{f}(x)$ of spin-preserving and spin-flipping properties. Furthermore, $V_{gc}(x)$ and $V_{gs}(x)$ can be arbitrary too. \\
In next section we shall discuss how this factorization property enables one to operatively determine, through transport properties, the transmission  coefficients \begin{eqnarray}
T_{p}&=&|t_{p}|^2= \frac{1}{|(\mathbf{m}_p)_{22}|^2}  \label{Tp-gen}\\
T_{f}&=&|t_{f}|^2= |(\mathbf{m}_f)_{22}|^2 \label{Tf-gen}  
\end{eqnarray} 
related to spin-preserving and spin-flipping tunneling, respectively. \\

The scattering matrix approach utilized here is non-perturbative, and it naturally accounts for the tunneling amplitudes $\Gamma_\nu(x)$ ($\nu=p,f$) to all perturbative orders. However, it is maybe worth clarifying the origin of the factorization in terms of perturbative arguments as well. Neglecting for simplicity the charge and spin gates, one would  derive the scattering amplitude of each multi terminal transport process by linear combinations of average values of $\hat\rho_{\alpha \sigma}$, performed over the time Keldysh contour $K$~\cite{keldysh},
\begin{eqnarray}
 \langle \hat\rho_{\alpha \sigma} \rangle =   \left\langle \hat\rho_{\alpha \sigma}   e^{ -\frac{i}{\hbar} \int_K dt   \Psi_{i}^\dagger\, (H^{p}_{\rm tun}+H^{f}_{\rm tun})_{ij} \Psi_{j}^{}  } \right\rangle_0  \quad,\label{Z}
\end{eqnarray}
where $\hat\rho_{\alpha \sigma}$ is the electron chiral density (\ref{rho-def}), $\Psi_i$ is the $i$-th component of the four-component electron field $\Psi(x)$ defined at the beginning of Sec.\ref{sec-II-a},  $H^{p}_{\rm tun}$, $H^{f}_{\rm tun}$ are the $4 \times 4$ matrices (\ref{Hp-m}) and (\ref{Hf-m}), and $\langle \ldots \rangle_0$ denotes the Keldysh average over $H_0$ [see Eq.(\ref{H0-m})]. For simplicity of notation we have omitted space integration and space-time arguments, and we have assumed implicit summation over repeated indices. 
We now notice from Eqs.(\ref{Hp-m})-(\ref{Hf-m}) that the lack of commutation between {p}- and {f}-processes,
\begin{equation}
\label{non-comm-m}
\left[ H^{p}_{\rm tun} \, , \,  H^{f}_{\rm tun}\right] \neq 0\quad, 
\end{equation}
is due to the appearance of the $\tau_z$  matrix in Eq.(\ref{Hf-m}), which is necessary to ensure time-reversal symmetry of $\mathcal{H}^{f}_{\rm tun}$ though. Expanding perturbatively   the r.h.s. of Eq.(\ref{Z}) in powers of the tunneling amplitudes $\Gamma_p$ and $\Gamma_f$, a given perturbative order is characterized by a power  $N_p$ for $H^{p}_{\rm tun}$ and by a power $N_f$ for $H^{f}_{\rm tun}$. Because $H^{p}_{\rm tun}$ involves same spin and opposite chirality, while  $H^{f}_{\rm tun}$ involves same chirality and opposite spin, one can realize that the only non vanishing contributions to   $\langle \hat\rho_{\alpha \sigma} \rangle$ occur when the integers $N_p$ and $N_f$ are both even ($N_p=2 n_p$ and $N_f=2 n_f$). Importantly, despite Eq.(\ref{non-comm-m}), one has 
\begin{equation}
\left[ (H^{p}_{\rm tun})^{2n_p} \, , \, ( H^{f}_{\rm tun})^{2n_f} \right] = 0
\end{equation}
Effectively, order by order, each non vanishing contribution to $ \langle \hat\rho_{\alpha \sigma} \rangle$ obtained from $H^{p}_{\rm tun}$ and $H^{f}_{\rm tun}$ is equal (up to a sign that counts the number of exchanges between $H^{p}_{\rm tun}$ and $H^{f}_{\rm tun}$)  to the contribution one would obtain by replacing $\tau_z \rightarrow \tau_0$ in Eq.(\ref{Hf-m}), i.e. by replacing  $H^{f}_{\rm tun}$ with a matrix that  commutes with $H^{p}_{\rm tun}$. One thus obtains  factorized expressions for the non-vanishing scattering matrix entries. \\

We conclude this section by noticing that the scattering matrix (\ref{S-res}) is {\it not} symmetric, despite the Hamiltonian of the system is time-reversal invariant. The customary expression of Onsager relations~\cite{onsager} characterizing the scattering matrix entries, $S_{ij}(\mathbf{B})=S_{ji}(-\mathbf{B})$ with $\mathbf{B}$ denoting the external magnetic field, would imply that $\mathsf{S}$ is symmetric in the absence of magnetic field. This is indeed the case for systems where spin is a good quantum number, which thus appears as a mere degeneracy variable. However, spin-flipping process can occur even without breaking of time-reversal symmetry, as is the case for {f}-processes for helical edge states in a tunnel junction. In this case, time-reversal transformation  involves  a $i\sigma_y$ matrix acting on the spin sector, i.e. a sign change whenever spin-$\downarrow$ is flipped to a spin-$\uparrow$. In this case Onsager relations acquire   different expressions. In particular,  any entry of the scattering matrix that describes a process  involving an odd number of spin flips naturally carries an additional minus sign. This is the reason for the appearance of both symmetric and anti-symmetric terms in the scattering matrix (\ref{S-res}).

%%%%%%%%%%%%%%%%%%%%%%%%%%%%%%%%%%%%%%%%%%%%%%%%%%%%%%%%%%
%%%%%%%%%%%%%%%%%%%%%%%%%%%%%%%%%%%%%%%%%%%%%%%%%%%%%%%%%%
%%%%%%%%%%%%%%%%%%%%%%%%%%%%%%%%%%%%%%%%%%%%%%%%%%%%%%%%%%
\section{Effects of the factorization  on transconductance}
\label{sec-III}
\begin{figure} 
\centering
\includegraphics[width=8cm,clip]{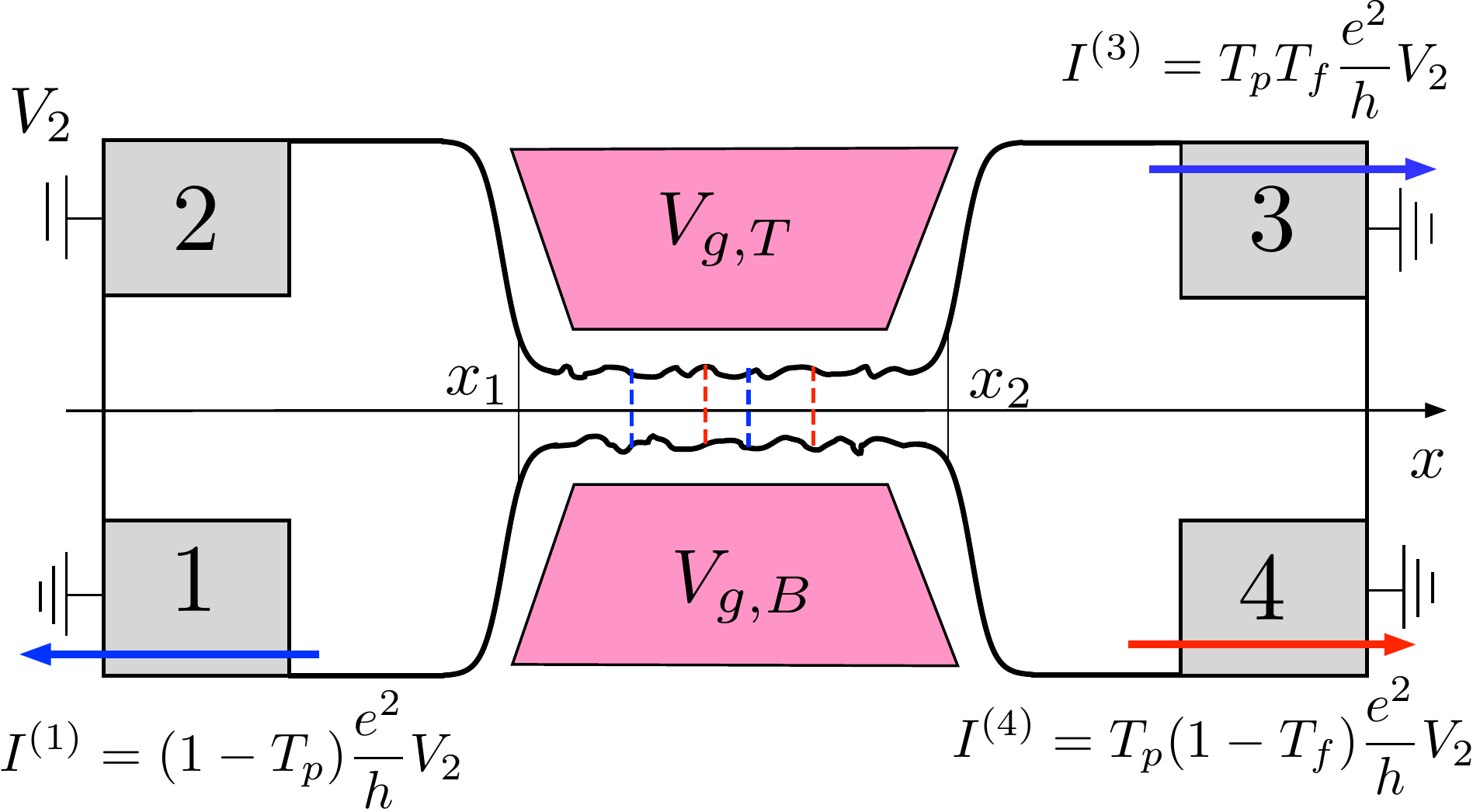}
\caption{(Color online) The operative way to extract the transmission coefficients $T_p$ and $T_f$ related to spin-preserving and spin-flipping tunneling, respectively. When a voltage bias $V_2$ is applied to terminal 2, the currents flowing in the other three terminals directly depend on $T_p$ and $T_f$ in a factorized form, so that $T_p$ and $T_f$ can be determined via Eqs.(\ref{Tp-exp})-(\ref{Tf-exp}). Furthermore, $T_p$ is controlled by the charge gate $V_{gc}=(V_{g,T}+V_{g,B})/2$, whereas $T_f$ is controlled by the spin gate $V_{gs}=(V_{g,T}-V_{g,B})/2$ }
\label{Fig-02-gedankenexp}
\end{figure}

We shall now present the results about transport through the setup, which are a direct consequence of the factorization (\ref{S-res}) of the scattering matrix entries. The currents operators related to the four terminals (denoted by $i=1 \ldots 4$ as in Fig.\ref{Fig-01}) are defined  as
\begin{equation}
\label{currents-def}
\begin{array}{lcl}
\hat{I}^{(1)}_c(x,t) &=& e v_F \left( \hat{\rho}_{R\downarrow}(x,t) -\hat{\rho}_{L\uparrow}(x,t)  \right) \\
\hat{I}^{(2)}_c(x,t) &=& e v_F \left( \hat{\rho}_{R\uparrow}(x,t) -\hat{\rho}_{L\downarrow}(x,t)  \right) \\
\hat{I}^{(3)}_c(x,t) &=& e v_F \left( \hat{\rho}_{L\downarrow}(x,t) -\hat{\rho}_{R\uparrow}(x,t)  \right) \\
\hat{I}^{(4)}_c(x,t) &=& e v_F \left( \hat{\rho}_{L\uparrow}(x,t) -\hat{\rho}_{R\downarrow}(x,t) \right)  
\end{array}
\end{equation}
and can then be evaluated   by substituting the solution (\ref{sol}) for a given tunnel junction into Eqs.(\ref{rho-def}) and (\ref{currents-def}), and integrating over energy $E$.
Notice that, in defining the currents in Eq.(\ref{currents-def}), we have chosen the customary convention of multi-terminal setups  that the current in a terminal is considered to be positive when it is incoming from that terminal to the scattering region~\cite{multi}.  We  recall that  in   multi-terminal transport  the  average  currents in the steady state are given by
\begin{equation} \label{av-cur}
I^{(i)} \doteq \langle  \hat{I}^{(i)}(x,t)\rangle \,= \frac{1}{e}  \sum_{j} \int_{-\infty}^{+\infty} \!dE \, {\rm G}_{ij} \, f_j
\end{equation} 
where $f_i=f_i(E)=[1+\exp((E-\mu_i)/k_B T)]^{-1}$ denotes the   Fermi function of the $i$-the reservoir, characterized by a temperature $k_B T$ and a chemical potential~$\mu_i$,   measured with respect to the equilibrium level $E_F$. In Eq.(\ref{av-cur}), ${\rm G}_{ij}$ denotes the entry of the conductance matrix~${\rm G}$, and describes the current flowing through  terminal $i$ as a consequence of a unit voltage bias applied to terminal $j$. The conductance matrix entries   can be expressed in terms of the Scattering matrix as ${\rm G}_{ij}=(e^2/h)(\delta_{ij}-|S_{ij}|^2)$.  The factorized form (\ref{S-res}) acquired by the scattering matrix    implies that the conductance matrix ${\rm G}$  reads
\begin{equation}
{\rm G}=\frac{e^2}{h} \left( \begin{array}{cccc}
1 & -R_{p}  &  -T_{p} R_{f}  & -T_{p} \, T_{f}  \\ & & & \\
%%%
-R_{p} & 1 & -T_{p} T_{f} & -T_{p} \, R_{f} \\ & & & \\
%%%
-T_{p} R_{f} & -T_{p} T_{f} & 1 & -R_{p}  \\ & & & \\
%%%
-T_{p} \, T_{f} & -T_{p} \,  R_{f} & -R_{p} & 1
\end{array}\right)
\label{G-res}
\end{equation}
At low temperatures the conductance matrix entry ${\rm G}_{ij}$ can be gained by measuring the current flowing in terminal $i$ in response to a voltage bias applied to only terminal~$j$ (transconductance), i.e.
\begin{equation}
{\rm G}_{ij}=\left. \frac{\partial I^{(i)}}{\partial V_j} \right|_{V_{l \neq j}=0} \quad.
\end{equation}
This provides and operative way to extract the transmission coefficients $T_p$ and $T_f$ through Eq.(\ref{G-res}).  
This is for instance carried out by   applying a voltage bias $V_2$ to terminal 2, while keeping all other chemical potentials to the equilibrium value, as illustrated in Fig.\ref{Fig-02-gedankenexp}. Then, the spin-preserving transmission coefficient $T_p$ is obtained  as
\begin{equation} \label{Tp-exp}
\begin{array}{lcl}
T_p &=& \displaystyle 1-\frac{h}{e^2} |{\rm G}_{12}| \\
&=& \displaystyle  1-\frac{h}{e^2} \left. \left| \frac{\partial I^{(1)}}{\partial V_2} \right| \right|_{V_{l \neq 2}=0}
\end{array} \quad,
\end{equation}
whereas and spin-flipping transmission coefficient  $T_f$ is
\begin{equation} \label{Tf-exp}
\begin{array}{lcl}
T_f &=& \displaystyle \frac{|{\rm G}_{32}|}{\frac{e^2}{h}- |{\rm G}_{12}|}=   \\ & &  \\
&=& \displaystyle \left.   \frac{\left| \frac{\partial I^{(3)}}{\partial V_2}   \right|}{\frac{e^2}{h}- \left|  \frac{\partial I^{(1)}  }{\partial V_2}   \right| }  \right|_{V_{l \neq 2}=0} \quad.
\end{array} 
\end{equation}
Similar equivalent methods, based on biasing another terminal and  measuring  currents in other appropriate terminals, can be read off from the structure of the ${\rm G}$ matrix~(\ref{G-res}), and can be used as cross checks for $T_p$ and $T_f$.
Importantly, because $T_{p}$ and $T_{f}$ exhibit different dependences
\begin{eqnarray}
T_{p}&=&T_{p}(\Gamma_{p};E_F-eV_{gc}) \\
T_{f}&=&T_{f}(\Gamma_{f};eV_{gs}) \quad,
\end{eqnarray}
they are controlled independently by the charge gate voltage $V_{gc}=(V_{g,T}+V_{g,B})/2$ and the spin gate voltage $V_{gs}=(V_{g,T}-V_{g,B})/2$, respectively.\\

We conclude this section by a remark concerning the operative method described in Fig.\ref{Fig-02-gedankenexp}. When a voltage bias is applied to terminal 2, the helical nature of the edge states implies that no current could be found in terminal 4 if only spin-preserving tunneling occurred in the junction. Thus, the very observation of  a current   in terminal 4 is a signature of the presence of spin-flipping processes. However, because {f}-processes interplay with {p}-processes too, the actual value of such current depends {\it also} on the latter, in a a-priori non-trivial way.   It is because of the factorization property proved here that such current appears to be simply proportional to $T_p(1-T_f)$, thereby enabling to extract the transmission coefficient~$T_f$ through Eq.(\ref{Tf-exp}).
%%%%%%%%%%%%%%%%%%%%%%%%%%%%%%%%%%%%%%%%%%%%%%%%%%%%
%%%%%%%%%%%%%%%%%%%%%%%%%%%%%%%%%%%%%%%%%%%%%%%%%%%%
%%%%%%%%%%%%%%%%%%%%%%%%%%%%%%%%%%%%%%%%%%%%%%%%%%%%%%%%%%%%%%%%%%%%%%%%%%%% 
%%%%%%%%%%%%%%%%%%%%%%%%%%%%%%%%%%%%%%%%%%%%%%%%%%%%%%%%%%%%%%%%%%%%%%%%%%%% 
%%%%%%%%%%%%%%%%%%%%%%%%%%%%%%%%%%%%%%%%%%%%%%%%%%%%%%%%%%%%%%%%%%%%%%%%%%%% 
\section{Explicit results for special cases}
\label{sec-IV}
The factorization property (\ref{S-res}) holds  for an  arbitrary profile of the tunnel junction parameters.    Studying the behavior of $T_p$ and $T_f$ with varying in all possible ways the  profiles $\Gamma_{p}(x), \Gamma_{f}(x)$ of the tunneling amplitudes and of the potentials $V_{gc}(x), V_{gs}(x)$ deserves a detailed analysis that goes beyond the purpose of the present paper. Nevertheless, in order to  show the potential of the result  found above,  in this section we  explicitly discuss some effects arising from the internal structure of a tunnel junction. We shall start in Sec.\ref{sec-IV-a} by considering the case  where the   parameters  $\Gamma_{p}(x)$, $\Gamma_{f}(x)$, $V_{gc}(x)$ and $V_{gs}(x)$ have a constant profile along the length $L$ of the junction. Such seemingly simplified model of the tunnel junction provides in fact quite useful physical insights. In the first instance it clarifies the different role of the {p}- and {f}-tunneling processes, and thereby the physical origin of the general property that $T_p$ depends on the energy $E$ and can be tuned  by $V_{gc}$, whereas $T_f$ is independent of the energy and can be tuned  by $V_{gs}$. Secondly, this case allows to  account for the effects of the finite length $L$ of the junction on both $T_p$ and $T_f$, which cannot be described by the conventional DT model of the tunnel region. \\ Then, in Sec.\ref{sec-IV-b} we show how  the constant-profile case can be   exploited to construct a realistic model of an actual tunnel junction with arbitrarily varying parameters. In Sec.\ref{sec-IV-c} we analyze the effect of a smooth variation of the absolute value $|\Gamma_{p}(x)|$ and  $|\Gamma_{f}(x)|$ of the tunneling amplitudes, whereas in   Sec.\ref{sec-IV-d} we analyze the effect of phase fluctuations $\phi_p(x)$ and $\phi_f(x)$. Finally, in Sec.\ref{sec-IV-DT} we   discuss the DT limit $L\rightarrow 0$, and to show why such widely used model subtly hides the factorization properties. 
%%%%%%%%%%%%%%%%%%%%%%%%%%%%%%%%%%%%%%%%%%%%%%%%%%%%%%%%%%%%%%
%%%%%%%%%%%%%%%%%%%%%%%%%%%%%%%%%%%%%%%%%%%%%%%%%%%%%%%%%%%%%%
%%%%%%%%%%%%%%%%%%%%%%%%%%%%%%%%%%%%%%%%%%%%%%%%%%%%%%%%%%%%%%
\subsection{The case of a constant profile}
\label{sec-IV-a}
\noindent Let us now consider constant profiles along the tunnel junction
\begin{equation}
\label{prof-const}
\Gamma_{\nu}(x)= \left\{ 
\begin{array}{lll}
 \, \Gamma_{\nu}  & \mbox{for } \, \,  x_1 < x < x_2 &\\
 &  &  \hspace{1cm} \nu=p,f \\
 0 & \mbox{otherwise} &  \\
\end{array} \right.
\end{equation}
and
\begin{equation}
\label{prof-const-V}
V_{gc(s)}(x)= \left\{ 
\begin{array}{lll}
 \,V_{gc(s)}  & \mbox{for } \, \,  x_1 < x < x_2 &\\
 &  &  \\
 0 & \mbox{otherwise} &  \\
\end{array} \right.
\end{equation}
Under the assumption (\ref{prof-const})-(\ref{prof-const-V}),  the $\mathbf{m}_f$-matrix (\ref{mf-def}) becomes
\begin{eqnarray}
\mathbf{m}_f &=& {\rm T} \, e^{  -i \int_{x_1}^{x_2}  \,\boldsymbol\sigma \cdot  \mathbf{b}_f    \, dx^\prime}  =      e^{ -i   \, (\boldsymbol\sigma \cdot  \mathbf{b}_f ) L }=\nonumber \\
&=&\sigma_0 \cos(\tilde{k}_{f} L) -i  \boldsymbol\sigma \cdot  \mathbf{b}_f \sin(\tilde{k}_{f} L)/ \tilde{k}_{f}
\label{u-const}
\end{eqnarray}
with
\begin{equation} \label{tilde-kf-def}
\tilde{k}_{f} = \sqrt{|\Gamma_{f}|^2+(e V_{gs})^2} \, /\hbar v_F \quad,
\end{equation}
whereas the $\mathbf{m}_p$-matrix (\ref{mp-def}) is given by 
\begin{eqnarray}
\lefteqn{ \mathbf{m}_p=   e^{-i \tau_z k_E x_2} \, ({\rm T} \, e^{-i \int_{x_1}^{x_2} \, \boldsymbol\tau \cdot \mathbf{b}_{p,E}  \, dx^\prime} ) e^{i \tau_z k_E x_1}=} & &  \nonumber \\
&=&  \, e^{-i \tau_z k_E x_2}    e^{-i   (\boldsymbol\tau \cdot \mathbf{b}_{p,E} ) \,L}   e^{i \tau_z k_E x_1} 
 =    \nonumber \\ & & \nonumber \\
&=&   \left\{ \begin{array}{l}
e^{-i \tau_z k_E x_2}    \left(\tau_0 \cos(\tilde{k}_E L)\,  -i  \boldsymbol\tau \cdot \mathbf{b}_{p,E}  \frac{\sin(\tilde{k}_E L) }{\tilde{k}_E} \right)  \cdot   \,  \\ \\ \hspace{1cm} \cdot \,  e^{+i \tau_z k_E x_1} \hspace{1.5cm} \mbox{for} \, \,  |E-eV_{gc}|>|\Gamma_{p}| \\ \\ \\
e^{-i \tau_z k_E x_2}  \left(\tau_0\cosh(\tilde{q}_E L) \,  - i \boldsymbol\tau \cdot \mathbf{b}_{p,E} \frac{\sinh(\tilde{q}_E L) }{ \tilde{q}_E} \right) \cdot   \, \\ \\
\hspace{1cm} \cdot \, e^{+i \tau_z k_E x_1} \hspace{1.5 cm} \mbox{for} \, \, |E-eV_{gc}|<|\Gamma_{p}| \end{array}\right.
\label{v-const}
\end{eqnarray}
where we have denoted

\begin{eqnarray}\label{tildekE_def}
\begin{array}{lcl} \tilde{k}_E = \frac{\sqrt{(E-eV_{gc})^2-|\Gamma_{p}|^2}}{\hbar v_F} & \mbox{for} & |E-e V_{gc}|>|\Gamma_{p}| \\ & & \\ \tilde{q}_E =\frac{\sqrt{|\Gamma_{p}|^2-(E-eV_{gc})^2}}{\hbar v_F}  &  \mbox{for} & |E-e V_{gc}|<|\Gamma_{p}|\end{array}  
\end{eqnarray}
The transmission coefficient related to spin-preserving tunneling is  obtained from the $\mathbf{m}_p$ matrix (\ref{v-const}) through Eq.(\ref{Tp-gen}), and reads
\begin{eqnarray}
\lefteqn{T_{p}(E)= } & & \nonumber  \\
&=& \left\{ \begin{array}{lll}  \displaystyle \frac{1}{1+\frac{|\Gamma_{p}|^2}{(\hbar  v_F \tilde{k}_E)^2}\, \sin^2{(\tilde{k}_E \, L) } } & \mbox{if} & |E-e V_{gc}|>|\Gamma_{p}|
\\ & & \label{Tp-theta} \\
\displaystyle \frac{1}{1+\frac{|\Gamma_{p}|^2}{(\hbar  v_F \tilde{q}_E)^2}\, \sinh^2{(\tilde{q}_E \, L) } } & \mbox{if} & |E-e V_{gc}|<|\Gamma_{p}|
\end{array} \right.   \label{Tp-const} 
\end{eqnarray}
whereas the transmission coefficient related to  spin-flipping tunneling is   easily obtained from the $\mathbf{m}_f$ matrix~(\ref{u-const}) through Eq.(\ref{Tf-gen}), and reads 
\begin{equation}
\label{Tf-theta} 
T_{f} = 1-  \frac{|\Gamma_{f}|^2}{(\hbar v_F \tilde{k}_{f})^2}  \sin^2[\tilde{k}_{f} L ]    \quad.  
\end{equation} 
Equations (\ref{Tp-const}) and (\ref{Tf-theta}), directly involve  $\Gamma_p$, $V_{gc}$ and $\Gamma_f$, $V_{gs}$ respectively, and allow to identify the physical effect of these terms. \\
%%%%%%%%%%%%%%%%%%%%%%%%%%%%%%%%%%%%%%%%%%%%%%%%%%%%
\begin{figure} 
\centering
\includegraphics[width=7cm,clip]{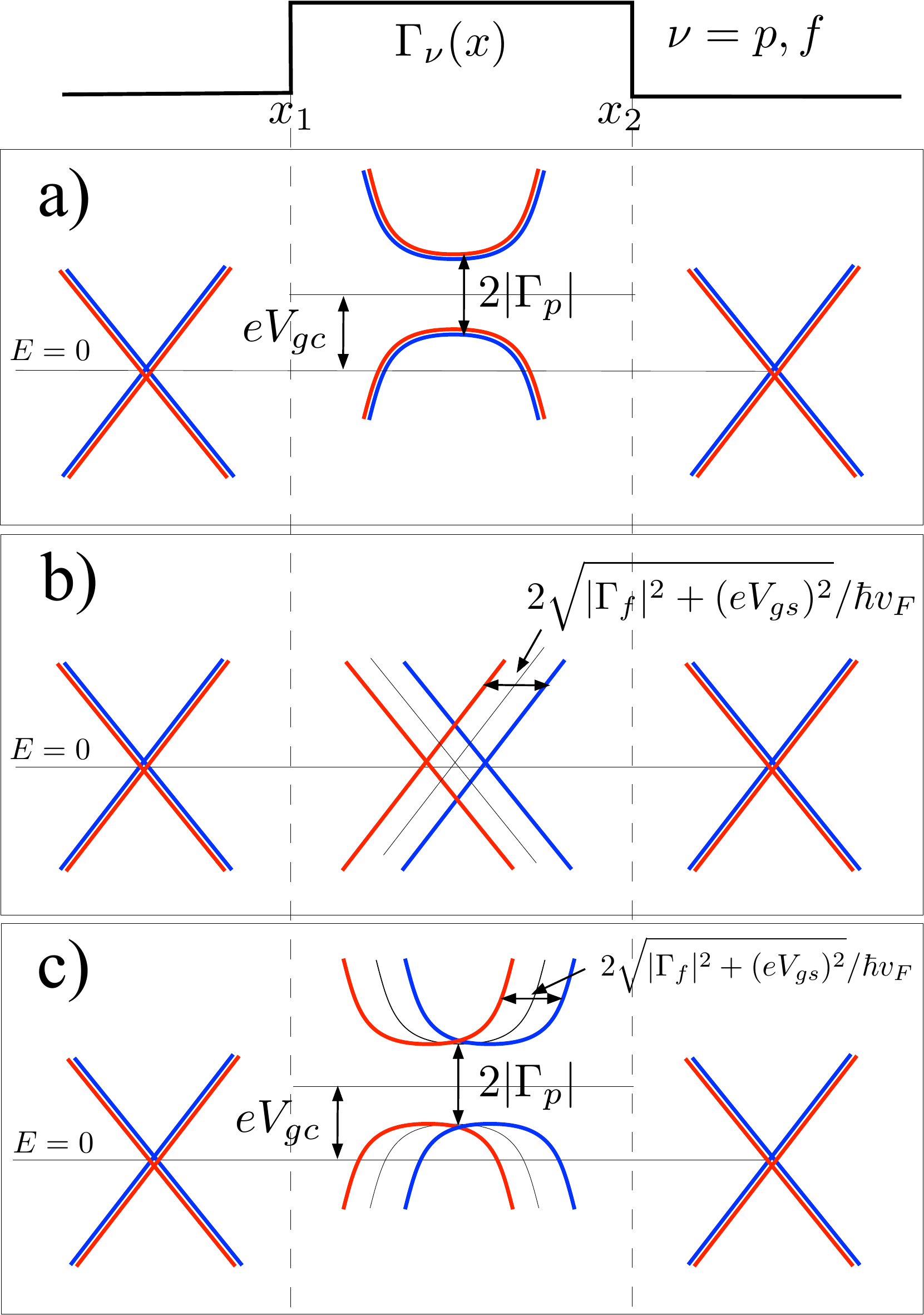}
\caption{(Color online) Effects of the spin-preserving and spin-flipping tunneling processes on the electronic spectrum of the edge states in the tunnel junction with profile (\ref{prof-const})-(\ref{prof-const-V}). a)  the spin preserving tunneling $\Gamma_p$ tends to induce a gap in the spectrum (which in fact amounts to a reduction of the transmission, due to the finite length of the junction), whereas  the charge bias $V_{gc}$ shifts vertically the spectrum    with respect to  the Fermi level.  b) the spin flipping tunneling $\Gamma_f$ lifts the degeneracy of spin-$\uparrow$ and spin-$\downarrow$ edge modes, inducing a mixing of the eigenstates spin-components in the tunnel junction as well as a mutual horizontal shift in the momenta. The spin gate $V_{gs}$ renormalizes such shift.  c) the effect of both tunneling terms.}
\label{Fig-03}
\end{figure}
%%%%%%%%%%%%%%%%%%%%%%%%%%%%%%%%%%%%%%%%%%%%%%%%%%%%

%%%%%%%%%%%%%%%%%%%%%%%%%%%%%%%%%%%%%%%%%%%%%%%%%%%%%%%%%%%%%%%%

\begin{figure} 
\centering
\includegraphics[width=7.5cm,clip]{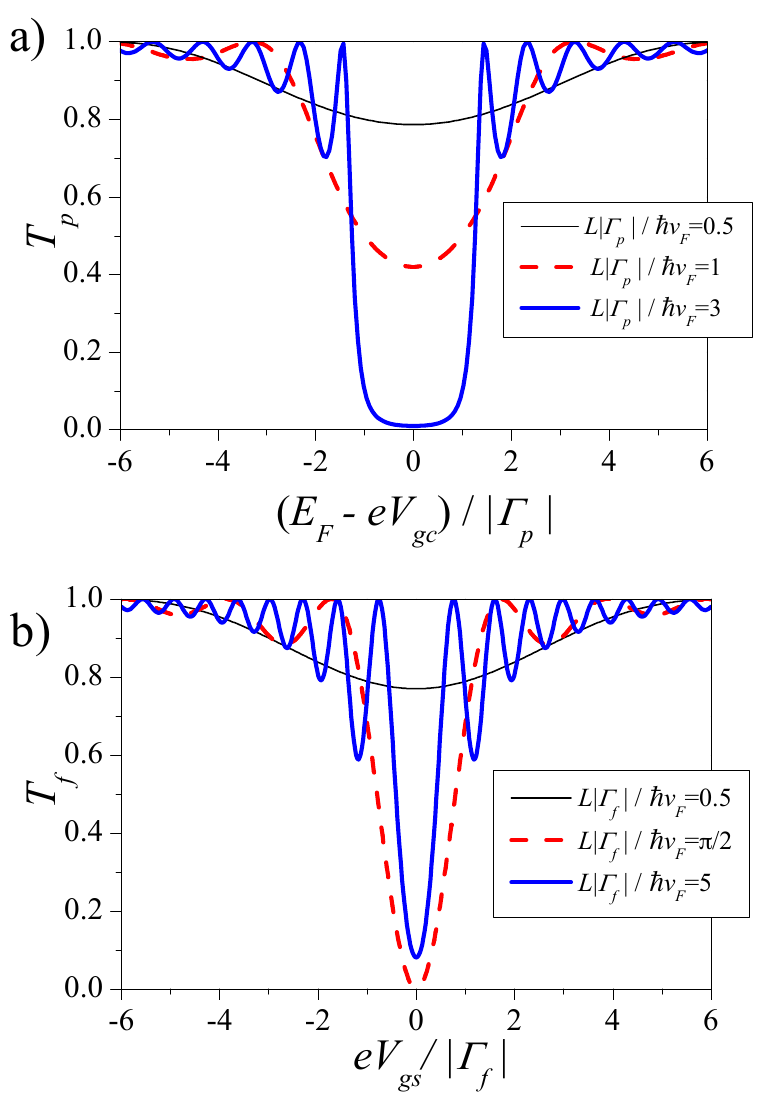}
\caption{(Color online) The case of a tunnel junction with a constant profile of the tunneling amplitude (\ref{prof-const}).    a) The  transmission coefficient $T_p$ due to spin-preserving tunneling (evaluated at the equilibrium Fermi level $E_F$) is plotted as a function of  
$E_F-eV_{gc}$ (with $V_{gc}$ denoting the charge gate voltage applied in the tunnel region), for different values of the length of the junction. In the energy range $|E_F-eV_{gc}|<|\Gamma_p|$, $T_p$ exhibits a minimum, which gradually becomes a `gap' with increasing the length of the junction [see also Fig.\ref{Fig-02-gedankenexp}a)], whereas in the energy range $|E_F-eV_{gc}|>|\Gamma_p|$ one can observe oscillations due to the interference of electronic waves backward scattered  at the ends of the tunnel junction.
The amplitude and frequency of the oscillations depend on the spin preserving strength of the tunnel junction $a_{p}=|\Gamma_{p}|L/\hbar v_{F}$: the oscillations increase in depth and frequency with increasing~$a_{p}$. 
  b) The  transmission coefficient $T_f$ due to spin-flipping tunneling  is plotted as a function of the spin gate voltage $V_{gs}$ for various values of the length of the junction. In this case no `gapped' energy range is observed [see also Fig.\ref{Fig-02-gedankenexp}b)]. The oscillations  are due to forward scattering interference and  increase in amplitude and frequency with increasing $a_{f}=|\Gamma_{f}|L/\hbar v_{F}$. }
\label{Fig-04}
\end{figure}

We start by analyzing the role of the  spin-preserving  tunneling. These processes would tend --per se--  to  create a gap in the electronic spectrum [see Fig.\ref{Fig-03}a)]~\cite{richter}. In fact, an actual gap would  be present   only for an infinitely long tunnel junction ($L\rightarrow \infty$), whereas in a realistic tunnel junction with a finite length $L$   two energy regimes can be identified. For $|E-eV_{g,c}|<|\Gamma_{p}|$ the electronic states in the tunnel junction region consist of evanescent waves in the longitudinal direction $x$, which decay as $\sim \exp[-\tilde{q}_E |x|]$, where $\tilde{q}_E$ is given in Eq.(\ref{tildekE_def}). In contrast, for $|E- eV_{g,c}|>|\Gamma_{p}|$ one has propagating waves, where the dispersion relation characterized by $\tilde{k}_E$ [see Eq.(\ref{tildekE_def})] is not linear though, due to the inter-edge coupling. This explains why $T_p$  depends on the energy~$E$. Notice that  the charge gate voltage $V_{gc}$ produces a vertical shift of the dispersion relation by changing the position of the Fermi level with respect to the Dirac point, so that $T_p$ is controlled by $V_{gc}$. 
These effects determine the  behavior of $T_{p}$, which is plotted in Fig.\ref{Fig-04}a)  at the equilibrium Fermi energy $E=E_F$ as a function of  $E_F-eV_{gc}$, for different values of the tunnel junction length $L$. The   minimum of the transmission coefficient at the Dirac point $E_F=eV_{gc}$ corresponds to the highest value $\tilde{q}_{E_F} L=|\Gamma_p| L/\hbar v_F$ of the evanescent wave  decay rate along the whole junction takes. While for short junction $|\Gamma_p| L/\hbar v_F<1$ the value of the minimum is finite, by increasing  the length $L$ of the junction and/or the tunnel coupling $\Gamma_p$, one observes a strong suppression of the $T_p$ minimum, which becomes a minigap as soon as $|\Gamma_p| L/\hbar v_F>1$.  \\

In contrast, the spin-flipping tunneling $\Gamma_f$ lifts the degeneracy of spin-$\uparrow$ and spin-$\downarrow$ energy bands, by introducing an equal and opposite horizontal shift by $\pm \tilde{k}_f$ in the momenta of the dispersion relation, where $\tilde{k}_f$ is given by Eq.(\ref{tilde-kf-def}). Such shift is {\it independent} of the energy $E$ of the incoming electron [see Fig.\ref{Fig-03}b)], which explains why $T_f$ is energy independent. Nevertheless, $\tilde{k}_f$ depends on $V_{gs}$, so that $T_f$ can be controlled by the spin gate.  Importantly, although $|\Gamma_f|$ and $V_{gs}$ have the same effect on the dispersion relation, they have a quite different effect  on the eigenstates. Indeed, while $V_{gs}$ lifts the degeneracy by preserving the eigenstates, the coupling $\Gamma_f$ mixes spin-$\uparrow$ and $\downarrow$ states, and  in the tunnel junction the eigenstates are not characterized by a unique spin orientation. Due to the factorization property these  features hold  also when both spin-preserving and spin-flipping tunneling are present [see Fig.\ref{Fig-03}c)]. In Fig.\ref{Fig-04}b), the spin-flipping transmission coefficient  $T_{f}$ [see Eq.(\ref{Tf-theta})]  is plotted as a function of the spin gate voltage $V_{gs}$, for different values of the junction length.  \\

%%%%%%%%%%%%%%%%%%%%%%%%%%%%%%%%%%%%%%%%%%%%%%%%%%%%
%%%%%%%%%%%%%%%%%%%%%%%%%%%%%%%%%%%%%%%%%%%%%%%%%%%%

\begin{figure} 
\centering
\includegraphics[width=7.5cm,clip]{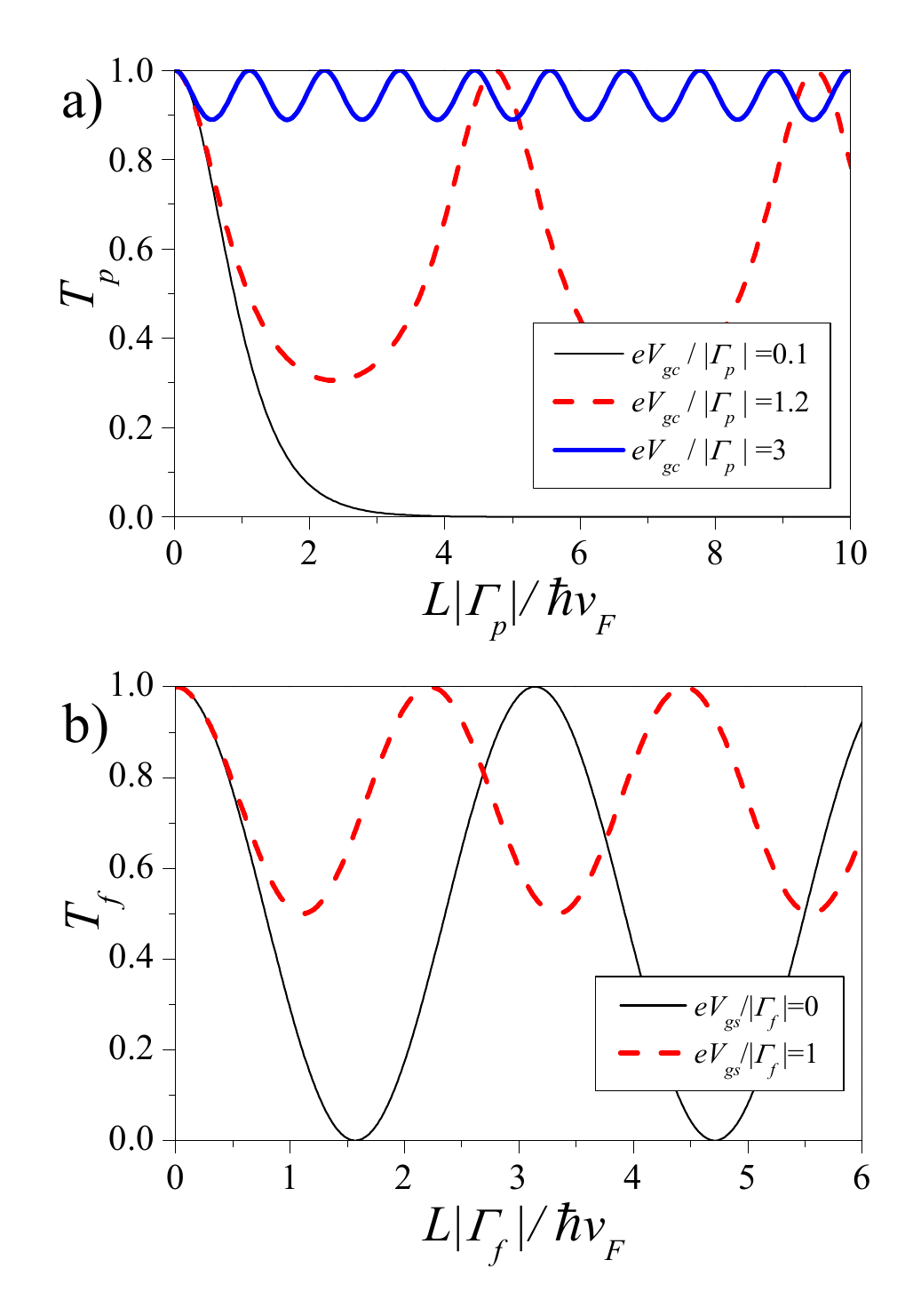}
\caption{(Color online) The behavior of the transmission coefficients $T_p$ and $T_f$ as a function of the length $L$ of the tunnel junction is typically non-monotonous.  a) The spin preserving transmission coefficient $T_p$ (evaluated at the energy $E_F=0$) is plotted as a function of $L |\Gamma_p|/\hbar v_F$, for various charge gate voltages. While for $|eV_{g,c}|>|\Gamma_{p}|$ it oscillates with $L$, in the regime $|eV_{g,c}|<|\Gamma_{p}|$ it exponentially decreases.    b) $T_f$ is plotted as a function of the length of the junction for various voltages $V_{gs}$. The behavior is always oscillatory. For  particular values of the junction length $L |\Gamma_f|/\hbar v_F=(m+1/2)\pi$ (at $eV_{gs}=0$) the coefficient $T_f$   vanishes,    leading to a complete spin flip.}
\label{Fig-05}
\end{figure}

We notice that both $T_p$ and $T_f$ exhibit oscillations, which are both a signature of electron quantum interference, although the origin is different.  
The oscillations of $T_p$ [see Fig.\ref{Fig-04}a) and Fig.\ref{Fig-05}a)] originate from  the spin-preserving tunneling that changes the group velocity. It is therefore an interference induced by backward-scattering at the two ends of the tunnel junctions, where the phase difference of the interfering waves is controlled by the charge gate voltage $V_{gc}$. The amplitude and frequency of these oscillations depend on the spin-preserving `strength' of the tunnel junction, i.e. on the dimensionless junction parameter $a_p=|\Gamma_p|L/ \hbar v_F$, combining length $L$ and tunneling amplitude $|\Gamma_p|$. Indeed the minima occur at energies   $|E_F-eV_{gc}| \simeq |\Gamma_p| \sqrt{1+ (m+1/2)^2 \, (\pi/a_p)^2}$, and their related values are approximately $(1+(a_p/\pi)^2 /(m+1/2)^2)^{-1}$, with  $m \in \mathbb{Z}$. This is in agreement with the results of Ref.[\onlinecite{citro-sassetti}], where only spin-preserving tunneling inside the tunnel junction was considered. In practice, the oscillations increase in depth and frequency with increasing strength $a_p$.  This implies that, while in the regime $|E_F -eV_{g,c}|<|\Gamma_{p}|$ the transmission coefficient $T_p$  decreases  as a function of the tunnel junction strength $a_p$, in  the regime $|E_F- eV_{g,c}|>|\Gamma_{p}|$ it oscillates with~$L$, as   shown in Fig.\ref{Fig-05}a), where   $T_p$ at $E_F=0$ is plotted as a function of $a_p$. The two different dependences can be accessed by a tuning the charge gate voltage $V_{gc}$.\\

%%%%%%%%%%%%%%%%%%%%%%%%%%%%%%%%%%%%%%%%%%%%%%%%%%%%
%%%%%%%%%%%%%%%%%%%%%%%%%%%%%%%%%%%%%%%%%%%%%%%%%%%%

In contrast, the oscillations of $T_f$ [see Fig.\ref{Fig-04}b) and Fig.\ref{Fig-05}b)] originate from f-tunneling processes, which couple electronic waves with the same chirality. To illustrate this effect, let us imagine that a   right-moving electron wave with spin-$\uparrow$ is injected from terminal 2. Due to f-tunneling process,  at the left end of the tunnel junction the electronic wave is split into two components, both propagating rightwards  but with opposite spin orientation. At the right end of the junction another spin-flipping tunneling process may flip the spin-$\downarrow$ component back to spin-$\uparrow$, inducing interference with the transmitted wave. It is therefore a forward-scattering interference, where the phase difference $\tilde{k}_{f}L$ accumulated along the junction is determined by the spin-gate $V_{gs}$. The  amplitude and frequency of these oscillations   depend  on the spin-flipping `strength' of the tunnel junction, i.e. on the dimensionless parameter $a_f=|\Gamma_f|L/\hbar v_F$. Indeed the minima occur at spin gate voltage values $eV_{gs} \simeq |\Gamma_f|\sqrt{\pi^2 (m+1/2)^2/a_f^2-1}$, and approximately take values $1-(a_f/\pi)^2/(m+1/2)^2$, with $m \in \mathbb{Z}$. 
In turn, this interference effect also implies a non-monotonous dependence of $T_f$ on the length of the junction, as shown  in Fig. \ref{Fig-05}b).   
With varying the value of $V_{gs}$, one can thus control  the percentage of the transmitted current that flows to terminal 4 with  spin-$\downarrow$ with respect to the current flowing to terminal 3 with a spin-$\uparrow$. Notice that, for particular values $a_f 
=(m+1/2)\pi$ of the tunnel junction spin-flipping `strength' and for $eV_{gs}=0$, the transmission $T_f$ can even vanish, leading to a complete spin-flip. This proves that spin-flipping tunneling processes have dramatic impact on transport, where the helical nature of the edge states can be exploited to realize tunable spin polarizers for spintronics applications. \\

In sec.\ref{sec-IV-c} we shall discuss how the oscillations of $T_p$ and $T_f$ change when the sharp transition from  vanishing tunneling amplitudes $\Gamma_p$ and $\Gamma_f$ outside the junction to a constant tunneling amplitude inside the junction  [see Eq.(\ref{prof-const})]   is replaced by a smoother profile.

%%%%%%%%%%%%%%%%%%%%%%%%%%%%%%%%%%%%%%%%%%%%%%%%%%%%%%%%%%%%%%%%%%%%%%%%%%%% 
%%%%%%%%%%%%%%%%%%%%%%%%%%%%%%%%%%%%%%%%%%%%%%%%%%%%%%%%%%%%%%%%%%%%%%%%%%%% 
%%%%%%%%%%%%%%%%%%%%%%%%%%%%%%%%%%%%%%%%%%%%%%%%%%%%%%%%%%%%%%%%%%%%%%%%%%%% 
\subsection{Generalization to an arbitrary profile}
\label{sec-IV-b}
\begin{figure} 
\centering
\includegraphics[width=7.5cm,clip]{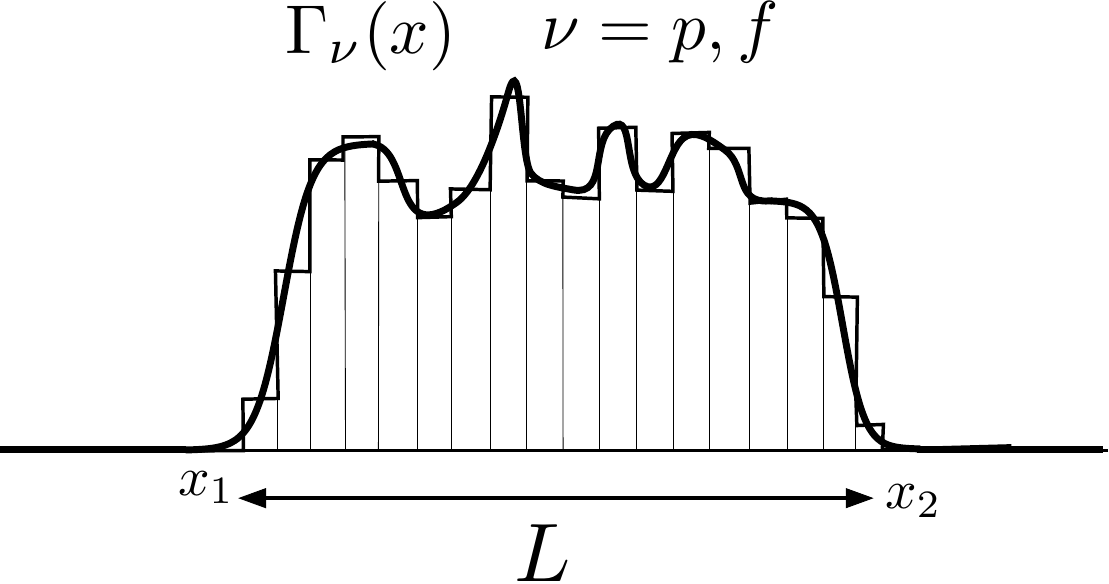}   
\caption{Sketch of the tunneling amplitude profile $\Gamma_\nu(x)$ in the tunnel junction region $x_1 \le x \le x_2$ ($\nu=p$ for spin-preserving and $\nu=f$ for spin-flipping tunneling processes, respectively). An  arbitrary profile  can be approximated with a stair-case profile, and the transfer matrix is easily obtained as a product of the constant profile transfer matrices [see Eq.(\ref{m-as-prod})].}
\label{Fig-06-profile}
\end{figure}

For an arbitrary profile of tunneling couplings $\Gamma_{p}(x)$ and $\Gamma_{f}(x)$ (see thick line in Fig.\ref{Fig-06-profile}),  the factorization property (\ref{M-factorized}) of the transfer matrix implies that one can compute $\mathbf{m}_p$ and $\mathbf{m}_f$ separately, and that the transmission coefficients $T_p$ and $T_f$ are then  evaluated through Eqs.(\ref{Tp-gen})-(\ref{Tf-gen}). In order to compute $\mathbf{m}_p$ and $\mathbf{m}_f$, we notice that the profile can be approximated with the desired accuracy by a $N$-step stair-case profile (thin line of  Fig.\ref{Fig-06-profile}). Then, exploiting a general property of transfer matrices~\cite{ihn-book}, $\mathbf{m}_p$ is straightforwardly  obtained as the product of the $\mathbf{m}_p^{(n)}$ matrices related to each individual $n$-th stair step, characterized by a locally constant profile  (here $n=1,\ldots N$, from right to left). Similarly for $\mathbf{m}_f$, obtaining
\begin{equation}
\label{m-as-prod}
\mathbf{m}_\nu = \prod_{n=1}^{N} \mathbf{m}_\nu^{(n)} \hspace{0.5cm} \nu=p,f
\end{equation}
One can thus use the case of constant profile investigated in Sec.\ref{sec-IV-a} as building block to model a tunnel junction with arbitrary profiles for $\Gamma_p(x)$ $\Gamma_f(x)$, $V_{gc}(x)$ and $V_{gs}(x)$.  In the following sections we apply this general method to investigate the effects of smoothing length and phase fluctuations in the tunneling amplitudes $\Gamma_\nu(x)$. For simplicity we shall restrict the gate profiles $V_{gc}(x)$ and $V_{gs}(x)$ to a constant gate profile  (\ref{prof-const-V}).
%%%%%%%%%%%%%%%%%%%%%%%%%%%%%%%%%%%%%%%%%%%%%%%%%%%%
%%%%%%%%%%%%%%%%%%%%%%%%%%%%%%%%%%%%%%%%%%%%%%%%%%%%
%%%%%%%%%%%%%%%%%%%%%%%%%%%%%%%%%%%%%%%%%%%%%%%%%%%%%%%%%%%%%%%%%%%%%%%%%%%% 
%%%%%%%%%%%%%%%%%%%%%%%%%%%%%%%%%%%%%%%%%%%%%%%%%%%%%%%%%%%%%%%%%%%%%%%%%%%% 
%%%%%%%%%%%%%%%%%%%%%%%%%%%%%%%%%%%%%%%%%%%%%%%%%%%%%%%%%%%%%%%%%%%%%%%%%%%% 
\subsection{Effects of a finite smoothing length}
\label{sec-IV-c}

\begin{figure} 
\centering
\includegraphics[width=7.5cm,clip]{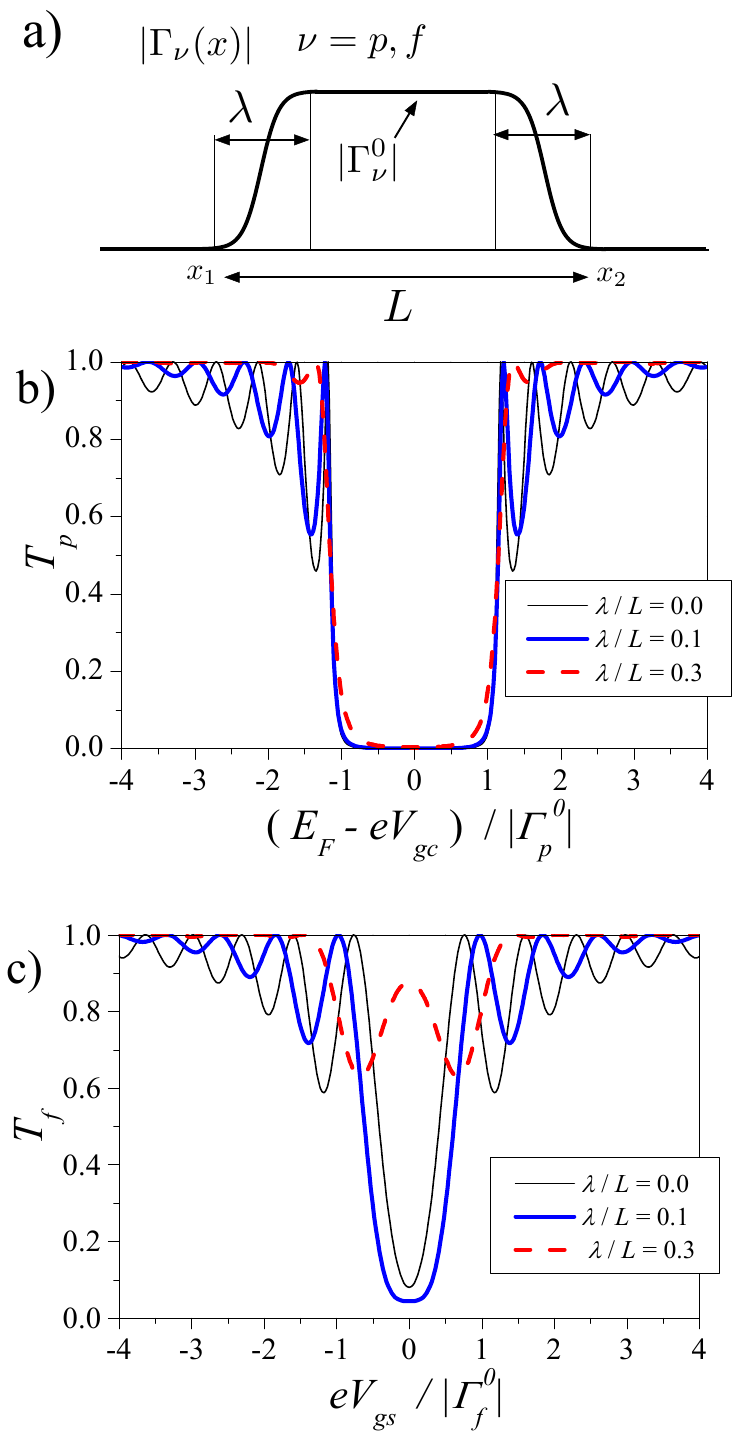} 
\caption{(Color online) a) Sketch of a tunnel amplitude profile $\Gamma_{\nu}(x)$, where the transition from a vanishing tunneling coupling to a bulk value $\Gamma_\nu^0$ occurs over a smoothing length scale $\lambda$ (here $\nu=p,f$ for spin-preserving and spin-flipping processes, respectively), where $L$ is the total length of the junction.    b) The effect of a smoothing length $\lambda$ on the spin preserving transmission coefficient: $T_p$ (evaluated at the equilibrium Fermi level $E_F$) is plotted as a function of $E_F-eV_{gc}$, for different values of the ratio  $\lambda/L$, for the case $L|\Gamma^{0}_{p}|/\hbar v_{F}=5$. While the mini gap region is essentially unaffected by $\lambda$, the amplitude and frequency of the oscillations is reduced  as $\lambda/L$ increases. They are still visible as long as one can define a longitudinal `bulk' of the tunnel junction with a constant value $\Gamma_0$, i.e. for $\lambda/L<0.25$.  c) The effect of a smoothing length $\lambda$ on the spin-flipping transmission coefficient:  $T_{f}$ is plotted as a function of $eV_{g,s}$, for the case $L|\Gamma^{0}_{f}|/\hbar v_{F}=5$. With increasing $\lambda$, the minimum at $V_{gs}=0$ exhibits  a non-monotonous behavior,  decreasing for $\lambda/L =0.1$, and then increasing and even turning into a local maximum for $\lambda/L=0.3$. Similarly to the oscillations of $T_p$, also for $T_f$ the amplitude of the oscillations is suppressed when the tunnel junction profile becomes very smooth.  }
\label{Fig-07-smoothing}
\end{figure}

We investigate here the case where the tunneling amplitudes $\Gamma_{\nu}(x)$ ($\nu=p,f$) change from a vanishing value (outside the tunnel junction)  to a `bulk' value $\Gamma_\nu^0$  over a {\it finite} smoothing length   $\lambda$, as shown in the profile depicted in Fig.\ref{Fig-07-smoothing}a). For simplicity, we shall consider the variation of the absolute value $|\Gamma_\nu(x)|$, and assume that the phase $\phi_\nu(x)$ remains constant. For a given value of the smoothing length $\lambda$, the actual profile is approximated by a stair-case profile  with $N_\lambda$  steps in the smoothing region $\lambda$, as   described in Sec.\ref{sec-IV-b}. The value of $N_\lambda$ is increased until convergence in the   transmission coefficients is reached. In practice, it turns out that a number $N_\lambda  \sim 20 \lambda/L$ of steps is sufficient to reach a convergence in the results, so that e.g. for $\lambda/L=0.2$ the correct result is obtained by using only $N_\lambda=4$. This proves that the method can be implemented with ordinary numerical routines in a treatable and fast way, thereby proving its usefulness  and flexibility in handling  arbitrary tunnel junction.   \\

%%%%%%%%%%%%%%%%%%%%%%%%%%%%%%%%%%%%%%%%%%%%%%%%%%%%
%%%%%%%%%%%%%%%%%%%%%%%%%%%%%%%%%%%%%%%%%%%%%%%%%%%%
The  results are plotted in Fig.\ref{Fig-07-smoothing}.  In particular, in Fig.\ref{Fig-07-smoothing}b)  the spin preserving   transmission coefficient $T_p$ (evaluated at the equilibrium Fermi energy $E_F$) is plotted as a function of $E_F-eV_{gc}$, for different values of the smoothing length $\lambda$, with $\lambda=0$ corresponding to the case of the constant profile discussed in \ref{sec-IV-a}. With increasing $\lambda$, the suppression of the transmission coefficient in the `sub-gap' region $|E_F-eV_{gc}|<|\Gamma_p^0|$  remains unaffected, whereas the visibility of the oscillations appearing in the `supra-gap' regime $|E_F-eV_{gc}|>|\Gamma_p^0|$   tends to be suppressed. The spin-flipping  transmission coefficient $T_f$, plotted in Fig.\ref{Fig-07-smoothing}c), exhibits a non-monotonous behavior of the $V_{gs}=0$ minimum by increasing $\lambda$: with respect to the case of constant profile (thin solid black line) it decreases for $\lambda/L =0.1$, whereas it increases and even turns into a local maximum for $\lambda/L=0.3$. Similarly to the oscillations of $T_p$, also for $T_f$ the amplitude of the oscillations is suppressed when the tunnel junction profile becomes very smooth ($\lambda \sim L/2$), like in a quantum point contact~\cite{martins}. This is due to an effective averaging  over various lengths of the backscattering processes causing the interference behavior. In contrast,   the oscillations are fairly visible as long as a `bulk' of the junction can be identified (i.e. for $\lambda< 0.25$). By combining etching and lithographic techniques this can easily be realized in QSHE systems based on HgTe/CdTe quantum wells~\cite{buhmann}. In this case, because $|\Gamma_\nu(x)|$ depends exponentially on the transversal width of the tunnel junction~\cite{zhou}, it is reasonable that $|\Gamma_\nu(x)| \simeq {\rm const}$ inside the junction, and  rapidly vanishing $|\Gamma_\nu(x)|$ at the ends of the tunnel region.

%%%%%%%%%%%%%%%%%%%%%%%%%%%%%%%%%%%%%%%%%%%%%%%%%%%%%%%%%%%%%%%%%%%%%%%%%%%% 
%%%%%%%%%%%%%%%%%%%%%%%%%%%%%%%%%%%%%%%%%%%%%%%%%%%%%%%%%%%%%%%%%%%%%%%%%%%% 
%%%%%%%%%%%%%%%%%%%%%%%%%%%%%%%%%%%%%%%%%%%%%%%%%%%%%%%%%%%%%%%%%%%%%%%%%%%% 
\subsection{Effect of phase fluctuations}
\label{sec-IV-d}
%%%%%%%%%%%%%%%%%%%%%%%%%%%%%%%%%%%%%%%%%%%%%%%%%%%%
\begin{figure} 
\centering 
\includegraphics[width=7.5cm,clip]{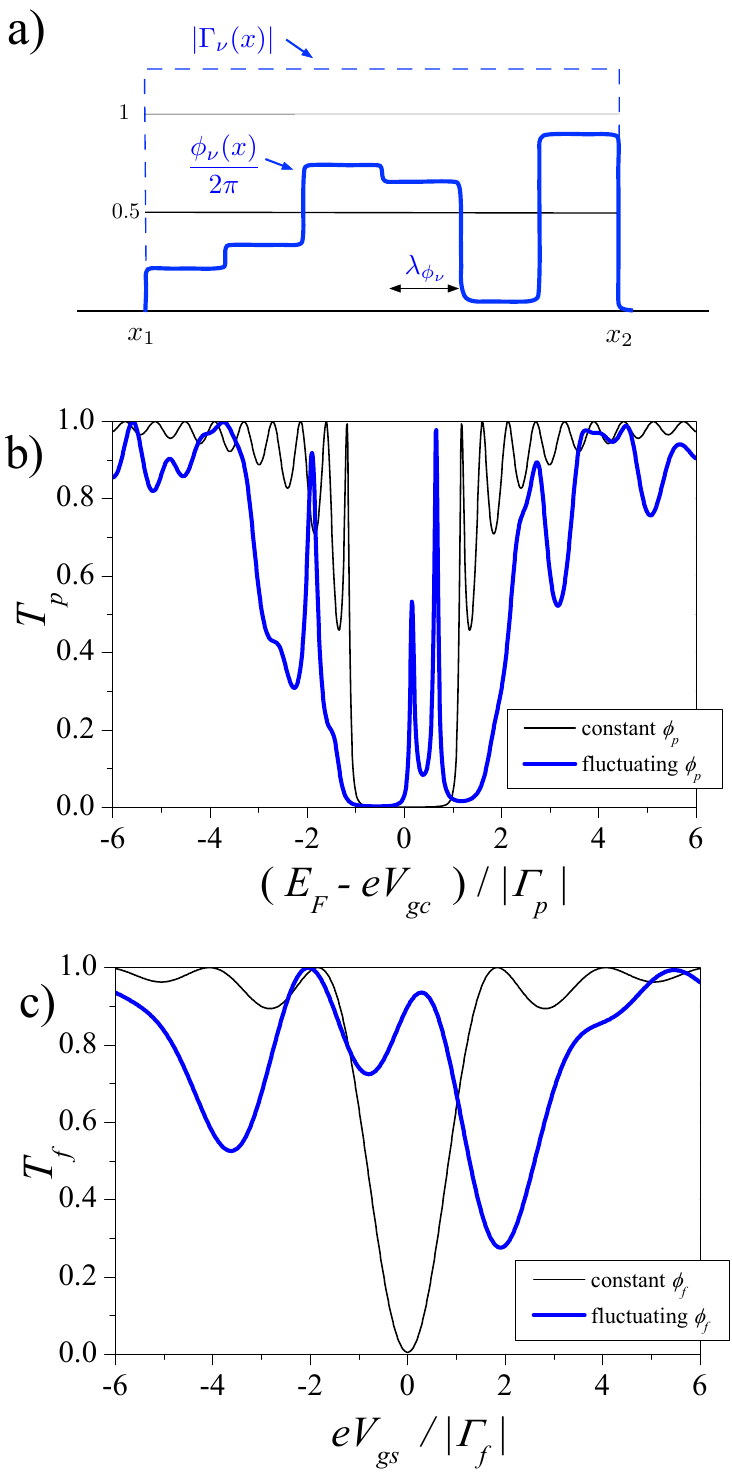} 
\caption{(Color online)   a) the phase $\phi_\nu$ of the tunneling amplitude $\Gamma_\nu(x)=|\Gamma_\nu(x)| \exp[i \phi_\nu(x)]$ is assumed to   fluctuate inside the junction over a typical length scale $\lambda_{\phi_\nu}$, whereas the absolute value $|\Gamma_\nu(x)|$ is assumed to be constant ($\nu=p$ for spin-preserving and $\nu=f$ for spin-flipping tunneling).  b) The spin-preserving transmission coefficient $T_{p}$ (evaluated at the equilibrium Fermi energy $E_F$) as a function of $E_F-eV_{gc}$,  for the case  $L|\Gamma_p|/\hbar v_F=5$. With respect to the case of constant $\phi_p$ (black thin line), the fluctuations of  $\phi_p$ induce resonance maxima appear inside the  
the `gap' region $|E_F-e V_{gc}|<|\Gamma_p|$, which in turn also broadens, whereas the oscillations   in the `supra-gap' region $|E_F-e V_{gc}|>|\Gamma_p|$ are enhanced (blue thick line). The curve becomes also asymmetric with respect to $E_F-e V_{gc}$.  
c) The spin-flipping transmission coefficient $T_{f}$ is plotted  as a function of $eV_{g,c}$, for the case $L|\Gamma_f|/\hbar v_F=1.5$: the fluctuations of the phase $\phi_f$ modify the oscillatory pattern with respect to the case of constant $\phi_f$ (black thin line), by increasing the  amplitude and making the plot asymmetric with respect to $V_{gs}=0$ (blue thick line).}
\label{Fig-08-phasejumps}
\end{figure}
%%%%%%%%%%%%%%%%%%%%%%%%%%%%%%%%%%%%%%%%%%%%%%%%%%%%

The tunneling amplitude is in general a complex function $\Gamma_\nu(x)=|\Gamma_\nu(x)| \exp[i \phi_\nu(x)]$, characterized by an absolute value and a phase. While the effect of variation of the absolute value $|\Gamma_\nu(x)|$ has been considered in Sec.\ref{sec-IV-a} and  Sec.\ref{sec-IV-c}, here we would like to focus on the role of the phase profile $\phi_\nu(x)$. To begin with, we observe that  for the case of a constant profile $\Gamma_\nu(x) \equiv \Gamma_\nu \exp[i\phi_\nu]$ [see Eq.(\ref{prof-const})]  the specific value of the phase $\phi_\nu$ is irrelevant, and only the absolute value $|\Gamma_\nu|$ matters in determining $T_p$ and $T_f$ [see Eqs.(\ref{Tp-theta})-(\ref{Tf-theta})]. While the assumption $|\Gamma(x)| \simeq \mbox{const}$ seems to be fairly reasonable in the central tunnel region, the situation may be different for the phase $\phi_\nu(x)$. One can expect that, especially in the presence of disorder, local potential fluctuations at each side of the tunnel junction may lead to random changes in the local Fermi wavevector $k_F$, affecting the electron phase $\sim \exp(i k_F x)$. At a given longitudinal position~$x$, the transversal overlap integral  determining the tunneling amplitude $\Gamma_\nu(x)$ may thus acquire phase fluctuations.  To discuss these effects we shall   consider a profile where the absolute value $|\Gamma_\nu(x)|$ remains constant, and the phase $\phi_\nu(x)$ fluctuates along the junction over a typical length scale $\lambda_{\phi_\nu}$, as shown in Fig.\ref{Fig-08-phasejumps}a). 
The effect on the transmission coefficients $T_p$ and $T_f$ are shown in panels b) and c). In particular, in panel b) $T_p$ is plotted  as a function of $E_F-eV_{gc}$. As compared to the case of a constant phase $\phi_\nu$ (black thin   line), the curve determined by phase fluctuations (blue thick line) exhibits various features: i) the appearance of some resonance maxima inside the `gapped' region $|E_F-e V_{gc}|<|\Gamma_p|$, whose  location  depends on $\lambda_{\phi_\nu}$ and on the typical deviation $\Delta\phi_p$ of the fluctuations around the average phase $\langle \phi_\nu \rangle $; ii) the broadening of the `sub-gap' region $|E_F-e V_{gc}|<|\Gamma_p|$;   iii) an enhancement of the amplitude of the oscillations in the `supra-gap' region $|E_F-e V_{gc}|>|\Gamma_p|$; iv) the symmetry of $T_p$   with respect to $E_F-eV_{gc}$ is lost.  The last two effects are particularly striking in the behavior of $T_f$ [see panel c)], where the amplitude of the oscillatory pattern is increased and $T_f$ is asymmetric in the spin gate bias $V_{gs}$.

%%%%%%%%%%%%%%%%%%%%%%%%%%%%%%%%%%%%%%%%%%%%%%%%%%%%%%%%%%%%%%%%%%%%%%%%%%%% 
%%%%%%%%%%%%%%%%%%%%%%%%%%%%%%%%%%%%%%%%%%%%%%%%%%%%%%%%%%%%%%%%%%%%%%%%%%%% 
%%%%%%%%%%%%%%%%%%%%%%%%%%%%%%%%%%%%%%%%%%%%%%%%%%%%%%%%%%%%%%%%%%%%%%%%%%%% 
\subsection{The delta-tunneling (DT) limit}
\label{sec-IV-DT} 
We now want to compare the case of finite length tunnel junction with the   widely used~\cite{teo,chamon_2009,strom_2009,trauz-recher,dolcini2012,dolcetto-sassetti,noise,akhmerov,dolcini2011,virtanen-recher,citro-romeo,citro-sassetti,liliana}
delta-tunneling model of a point-like constriction. Such model amounts to adopt sharply peaked   profiles for $\Gamma_{p}(x)$ and $\Gamma_{f}(x)$ in Eqs.(\ref{Hsp}) and (\ref{Hsf}),~ i.e. 
\begin{equation}
\Gamma_{\nu}(x)=2 \hbar v_F \, \gamma_{\nu}^{\mbox{\tiny \it DT}} \, \delta(x) \hspace{1cm} \nu=p,f \label{Gamma-DT} 
\end{equation}
where $\gamma_{p/f}^{\mbox{\tiny \it DT}} $ denote dimensionless  delta-tunneling amplitude  parameters. Solving the field equation of motion (\ref{eom-psi}) for the $\delta$-profile (\ref{Gamma-DT}), one can determine 
 the conductance matrix ${\rm G}^{\mbox{\tiny \it DT}}$ describing the  transmission coefficients between the four terminals, obtaining~\cite{dolcini2011} 
\begin{eqnarray}
{\rm G}^{\mbox{\tiny \it DT}}_{12} = {\rm G}^{\mbox{\tiny \it DT}}_{34} &= & -\frac{e^2}{h}\frac{4 \, \left | \gamma_{p}^{\mbox{\tiny \it DT}} \right |^2}{\left(1+\left | \gamma_{p}^{\mbox{\tiny \it DT}}\right | ^2+\left | \gamma_{f}^{\mbox{\tiny \it DT}}\right |^2\right)^2} \nonumber \\
{\rm G}^{\mbox{\tiny \it DT}}_{31} = {\rm G}^{\mbox{\tiny \it DT}}_{42}    &= &-\frac{e^2}{h} \frac{4 \left | \gamma_{f}^{\mbox{\tiny \it DT}}\right | ^2}{\left(1+\left | \gamma_{p}^{\mbox{\tiny \it DT}}\right | ^2+\left | \gamma_{f}^{\mbox{\tiny \it DT}}\right|^2\right)^2}  \label{T-DT} \\
{\rm G}^{\mbox{\tiny \it DT}}_{41} ={\rm G}^{\mbox{\tiny \it DT}}_{32}  &= & -\frac{e^2}{h}\frac{\left(1-\left | \gamma_{p}^{\mbox{\tiny \it DT}}\right |^2-\left | \gamma_{f}^{\mbox{\tiny \it DT}}\right | ^2\right)^2}{\left(1+\left |\gamma_{p}^{\mbox{\tiny \it DT}}\right |^2+\left | \gamma_{f}^{\mbox{\tiny \it DT}}\right |^2\right)^2} \nonumber \quad .  
\end{eqnarray}
Importantly, the coefficients (\ref{T-DT}) are {\it not} factorized into a spin-preserving and a spin-flipping contributions. This lack of factorization seems to  contradict  the result found above  for an arbitrary tunneling profile, since the DT model should be recovered from the finite length junction as the limit of $L \rightarrow 0$ of short length. 
To solve this seeming paradox one can proceed as follows. Instead of adopting the mathematical point-like tunneling profile (\ref{Gamma-DT}), one can follow a physically more correct procedure starting by a  model  where the tunnel junction has a finite length, and  taking the limit of vanishing length. This is for instance accomplished by assuming the constant profile described in Sec.\ref{sec-IV-a}, where $\Gamma_{\nu}(x)=\Gamma_{\nu} \,\theta(x-x_1) \theta(x_2-x)$, with $\Gamma_{\nu}=|\Gamma_{\nu}| e^{i\phi_{\nu}}$. Taking the limit of short tunnel junction  $L=x_2-x_1 \rightarrow 0$ and  $|\Gamma_{\nu}| \rightarrow \infty$,  with keeping the tunnel junction strengths $L|\Gamma_{\nu}| /\hbar v_F= \mbox{\rm const}$, one can operatively identify the expressions for the coefficients $\gamma_{p/f}^{\mbox{\tiny \it DT}}$ in terms of $L|\Gamma_{p}|/\hbar v_F$ and $L|\Gamma_{f}|/\hbar v_F$.   A lengthly but straightforward calculation leads to 
\begin{eqnarray}
 \gamma_{p}^{\mbox{\tiny \it DT}}  &=&  \frac{\sinh(\frac{L|\Gamma_{p}|}{\hbar v_F})  \, e^{i\phi_{p}}}{\cosh(\frac{L|\Gamma_{p}|}{\hbar v_F})+\cos(\frac{L|\Gamma_{f}|}{\hbar v_F})} \label{gammapDT-res} \\
 \gamma_{f}^{\mbox{\tiny \it DT}} &=&   \frac{\sin(\frac{L|\Gamma_{f}|}{\hbar v_F}) \, e^{i\phi_{f}}}{\cosh(\frac{L|\Gamma_{p}|}{\hbar v_F})+\cos(\frac{L|\Gamma_{f}|}{\hbar v_F})}\label{gammafDT-res} 
\end{eqnarray}
Equations (\ref{gammapDT-res})-(\ref{gammafDT-res}) show that the bare tunneling amplitudes $\gamma_{p/f}^{\mbox{\tiny \it DT}}$ used in the mathematical $\delta$-like profile (\ref{Gamma-DT}) actually depend  on {\it both} the physical spin-preserving and spin-flipping tunneling amplitudes $\Gamma_{p}$ and $\Gamma_{f}$ of the more realistic (i.e. narrow but finite) constriction model. 

\noindent  Physically, this seemingly surprising result can be understood as follows. Let us focus, for instance, on the spin-flipping channel:  A spin-flipping tunneling event  can be either direct, i.e. resulting from one single {f}-process, or indirect, i.e. `dressed' by  additional  tunneling events occuring along the junction. In particular, also an even number of {p}-processes can contribute to the spin-flipping tunneling, with a weight determined by the strength $a_p=|\Gamma_p|L/\hbar v_F$ of the spin-preserving tunnel coupling, which combines  the local tunneling amplitude $|\Gamma_p|$ and the length $L$ of the junction.  At first, one is tempted to think that  the DT limit, where the length $L$ of the tunnel junction vanishes, only involves direct tunneling events. However, because $|\Gamma_p|\rightarrow \infty$ and $|\Gamma_p| L$ is kept constant, dressing {p}-processes do matter if $|\Gamma_p|L/\hbar v_F \sim 1$, so that the parameter $\gamma_f^{\mbox{\tiny \it DT}}$ appearing in Eq.(\ref{Gamma-DT}) in fact describes the overall result of both the direct {f}-tunneling and all the dressing {p}-tunneling events.  Similarly for the other channel. 
At mathematical level, such effect originates from the fact that, when the size of the tunneling region becomes small, the wave function inside the constriction    stretches  and eventually becomes discontinuous in the limit $L \rightarrow 0$. The discontinuity is given by the integral $\int dx \, \Psi(x)$ over the tunneling region $x_1< x < x_2$. Thus, although the space `evolution' for $\Psi(x)$ is factorized into a product of {p}- and {f}- contributions [see Eq.(\ref{sol})], its integral is not, $\int dx \, (\mathbf{U}_{f}(x) \otimes \mathbf{U}_{p,E}(x)) \,  \, \, \neq \, \,  (\int dx \mathbf{U}_{f}(x)) \otimes  (\int dx\, \mathbf{U}_{p,E}(x))  $, so that $\int dx \, \Psi(x)$ depends  on both $\Gamma_{p}$ and $\Gamma_{f}$ in a non-trivial way. This integral is precisely what determines the parameters $\gamma_{p}^{\mbox{\tiny \it DT}}$ and $ \gamma_{f}^{\mbox{\tiny \it DT}}$ of the DT model (\ref{Gamma-DT}). Such singular behavior of the Dirac equation in the presence of $\delta$-like potential or tunneling profiles has been known since long~\cite{Dirac-and-delta-old}, and similar technical subtleties arise  also in other physical situations, such as the transport of chiral electrons in graphene through gapped regions~\cite{gomes-peres} and   tunneling through Majorana states appearing at the edges of superconductors~\cite{hou-refael}.  \\

Importantly, using Eqs.(\ref{gammapDT-res})-(\ref{gammafDT-res}) to re-express  the mathematical tunneling amplitude $\gamma_{p/f}^{\mbox{\tiny \it DT}}$ appearing in the DT model in terms of the  physical parameters $L\Gamma_{p}/\hbar v_F$ and $L\Gamma_{f}/\hbar v_F$, the conductance matrix entries (\ref{T-DT}) do  acquire the factorized form (\ref{G-res}), with spin-preserving and spin-flipping transmission coefficients 
\begin{eqnarray}
T_p^{\mbox{\tiny \it DT}} &= & \cosh^{-2} (L|\Gamma_{p}|/\hbar v_F)  \label{Tp-DT-1} \\
T_f^{\mbox{\tiny \it DT}} &=& \cos^2 (L|\Gamma_{f}|/\hbar v_F)  \quad, \label{Tf-DT-1} 
\end{eqnarray}
consistently  with the limit $L\rightarrow 0$ and $|\Gamma_{p/f}| \rightarrow \infty$ in Eqs.(\ref{Tp-const}) and (\ref{Tf-theta}). This proves that in the DT model the factorization is only seemingly lacking, and is hidden in the physically misleading parametrization in terms of~$\gamma_{p/f}^{\mbox{\tiny \it DT}}$. \\

This comparison allows one to identify the range of validity of the DT model (\ref{Gamma-DT}). The DT model is applicable when i) both the tunnel junction strength parameters are small (i.e. $L|\Gamma_\nu|/\hbar v_F \ll 1$, with $\nu=p,f$) and ii) when one is probing energy ranges smaller than $|\Gamma_\nu|$, i.e. when $|E_F-eV_{gc}| \ll |\Gamma_p|$ and $|V_{gs}| \ll |\Gamma_f|$. If the first condition is not fulfilled, than for each channel $\nu=p,f$ the related bare DT parameter  $\gamma_{\nu}^{\mbox{\tiny \it DT}}$ should actually be   `dressed' by contribution arising to higher-order processes of the other channel [see Eqs.(\ref{gammapDT-res})-(\ref{gammafDT-res})]. If the second condition is not fulfilled, the DT model cannot reproduce the features arising from the internal structure of the tunnel junction, such as the oscillatory pattern shown in Fig.\ref{Fig-04}. \\

Within the validity regime of the DT model, one can directly determine the tunneling parameters $|\Gamma_p|$ and $|\Gamma_f|$ to the measured transmission coefficients.  Indeed by performing transconductance measurements, one can extract [Eqs.(\ref{Tp-exp})-(\ref{Tf-exp})]  
\begin{eqnarray}
T_p^{\mbox{\tiny \it DT}} &=& 1- \frac{h}{e^2} |{\rm G}_{12}^{\mbox{\tiny \it DT}}|  \label{Tp-DT}  \\
T_f^{\mbox{\tiny \it DT}} &=&  \frac{|{\rm G}_{32}^{\mbox{\tiny \it DT}}|}{\frac{e^2}{h}- |{\rm G}_{12}^{\mbox{\tiny \it DT}}|}\quad .\label{Tf-DT}
\end{eqnarray}
Then, by using Eqs.(\ref{Tp-DT-1})-(\ref{Tf-DT-1}), one obtains
\begin{eqnarray}\label{gammap_gammaf}
|\Gamma_p| &=& \frac{\hbar v_F}{L} \mbox{\rm arctanh} (\sqrt{1-T_p^{\mbox{\tiny \it DT}}}) \\
|\Gamma_f| &=& \frac{\hbar v_F}{L} \arcsin (\sqrt{1-T_f^{\mbox{\tiny \it DT}}}) \quad.
\end{eqnarray}

%%%%%%%%%%%%%%%%%%%%%%%%%%%%%%%%%%%%%%%%%%%%%%%%%%%%%%%%%%
%%%%%%%%%%%%%%%%%%%%%%%%%%%%%%%%%%%%%%%%%%%%%%%%%%%%%%%%%%
%%%%%%%%%%%%%%%%%%%%%%%%%%%%%%%%%%%%%%%%%%%%%%%%%%%%%%%%%%
\section{Discussion and Conclusions}
\label{sec-V}
We have investigated a tunnel junction coupling the helical states flowing at the two edges of a 2D Quantum Spin Hall system (see Fig.\ref{Fig-01}), which can be realized by etching a constriction in a HgTe/CdTe or  InAs/GaSb quantum well. In such situation, electron tunneling  occurs through two types of time-reversal symmetric channels, namely spin-preserving ({p}) and spin-flipping ({f}) processes, making such system a bench  test for possible applications of helical edge states in spintronics. Indeed, due to the helical nature of the edge states, currents in the setup can in principle be switched from a terminal to another with either preserving or flipping the spin orientation. To this purpose, the crucial issue to is determine and control the transmission coefficients related to the two types of tunneling processes. This challenging task involves various difficulties, arising from the fact that the Hamiltonian terms describing these two processes do not commute, and  that a tunnel  region has a finite length and   a typically irregular and disordered profile, so that the tunneling amplitude of each of the two processes cannot be described by one single parameter but rather by a space-dependent profile.

We have demonstrated that there exist an operative way to separately extract the transmission coefficients $T_p$ and $T_f$, related to {p}- and {f}-processes, respectively. Indeed, despite the non-commutativity of the two tunneling terms, the analysis of the scattering matrix of the 4-terminal setup  has revealed that its entries are always factorized into two terms, one depending on the {p}-processes only and another one depending on the {f}-processes only. 
This factorization of the scattering matrix entries directly leads to the factorization of the Conductance matrix entries ${\rm G}_{ij}$, which determine the current flowing in  terminal $i$ when a voltage bias is applied to terminal $j$. It is thus possible to extract, via transconductance measurements, the transmission coefficients $T_p$ and $T_f$ related to these two processes (see Fig.\ref{Fig-02-gedankenexp}). Furthermore, by considering the presence of two electric gates across the junction, characterized by gate voltages $V_{g,T}$ and $V_{g,B}$,  we have shown that $T_p$ is controlled by the charge gate $V_{gc}=(V_{g,T}+V_{g,B})/2$ only, whereas $T_f$ is controlled by the spin gate $V_{gs}=(V_{g,T}-V_{g,B})/2$ only. \\

The factorization of the scattering matrix entries is seemingly lacking in the customary DT model for tunnel junction, and we have shown that it is  in fact subtly hidden in a misleading parametrization of the coupling constants of that model. In fact, we have proved that the factorization property  holds for an   arbitrary  profile of the tunneling amplitudes $\Gamma_\nu(x)$ ($\nu=p,f$). This enables one to go beyond the  DT model, and to investigate also the effects arising from the internal structure of the tunnel junction on the transmission    coefficients $T_p$ and $T_f$. We have first considered the effects of the finite length $L$ of the tunnel junction. In particular we have shown that $L$ determines the existence of two energy regimes on $T_p$. In the range $|E_F-eV_{gc}|<|\Gamma_p|$ the coefficient $T_p$ exhibits a minimum, which  is considerably suppressed and tends to acquire a gap-like feature when the length of the tunnel junction is increased [see Fig.\ref{Fig-04}a)], whereas in the range $|E_F-eV_{gc}|>|\Gamma_p|$ the coefficient $T_p$ exhibits an oscillatory behavior, with a period related to the length of the junction through the spin-preserving strength   $a_p=L|\Gamma_p|/\hbar v_F$. The effect of the finite length on the spin-flipping transmission coefficient $T_f$ is different. This is due to the fact that $\Gamma_f$ breaks the spin degeneracy of the helical states, without tendency to create a gap. Thus, $T_f$ exhibits a minimum as a function of the spin-gate potential $eV_{gs}$, which never becomes  a flat dip [see Fig.\ref{Fig-04}b)]. Oscillatory behavior is still present, similarly to the `supra-gap' region of $T_p$. The coefficients   $T_p$ and $T_f$ exhibit a non-monotonous behavior as a function of the length $L$ of the tunnel junction (see Fig.\ref{Fig-05}).\\

We have then investigated how the scenario changes when the profile of the tunneling amplitudes varies from a vanishing value outside the junction up to a constant value $\Gamma_\nu^0$ inside the junction   over a smoothing length $\lambda$   [see Fig.\ref{Fig-07-smoothing}a)]. As far as $T_p$ is concerned, the presence of the smoothing length $\lambda$ has a minor effect in the `sub-gap' region $|E-eV_{gc}|<|\Gamma_p|$, whereas   the amplitude of the oscillations occurring in the `supra-gap' region tends to be suppressed as $\lambda/L$ is increased [see Fig.\ref{Fig-07-smoothing}b)].  In contrast, as far as $T_f$ is concerned, the  minimum at $V_{gs}=0$ exhibits a non-monotonous behavior as a function of $\lambda$ and, for  $\lambda/L > 0.1$,   it turns into a maximum [see Fig.\ref{Fig-07-smoothing}c)]. As a whole, the amplitude of the oscillations are suppressed as $\lambda/L$ increases. However, the oscillations are still visible as long as a longitudinal `bulk' with a roughly constant $|\Gamma_\nu^0|$ can be identified, i.e. for $\lambda/L < 0.25$.
\\Then, we have investigated the role of fluctuations of the phase $\phi_\nu$ of the tunneling amplitude [see Fig.\ref{Fig-08-phasejumps}a)] $\Gamma_\nu(x)=|\Gamma_\nu(x)|\exp[i\phi_\nu(x)]$, arising from disorder in the tunnel junction. We have seen that, in the presence of random fluctuations of the spin-preserving tunneling amplitude phase $\phi_p$,  resonances appear in the `sub-gap' region $|E_F-eV_{gc}|<|\Gamma_p|$ of $T_p$, whereas the amplitude of the oscillations in the `supra-gap' region $|E_F-eV_{gc}|>|\Gamma_p|$ is also enhanced  [see Fig.\ref{Fig-08-phasejumps}b)]. Furthermore, $T_p$ is no longer symmetric with respect to $E_F-eV_{gc}$ in the presence of such fluctuations.
On the other hand, the  fluctuations of the phase~$\phi_f$ of the spin-flipping tunneling amplitude lead to an enhancement of the amplitude and location of the oscillations of $T_f$ as a function of the spin gate voltage $V_{gs}$~[see Fig.\ref{Fig-08-phasejumps}c)].\\

{\it Experimental conditions}. Let us now briefly discuss the experimental conditions to realize the setup. It is well known that, similarly to graphene, the helical edge states of QSHE cannot be confined simply by electrical gating, due to the linear Dirac-like spectrum and the Klein tunneling. Tunnel junctions in QSHE are thus typically realized by lateral etching of the quantum well, and lithographic techniques can be exploited to tailor arbitrary  shapes.  Once the etched constriction induces electron tunneling, lateral gates can be used to control it, as described above.    For a typical tunnel region of a width $W \sim 100 \, {\rm nm}$, one can estimate $|\Gamma_p| \sim 1.3 \, {\rm meV}$ and $|\Gamma_f| \sim 0.3 \, {\rm meV}$~\cite{richter,citro-sassetti,zhou}. Notice that $|\Gamma_f| \sim |\Gamma_p|/4$, i.e. $|\Gamma_f|$ is smaller, but not negligible with respect to $|\Gamma_p|$. These values, together with the length $L$ of the junction, determine the variation range for the gate voltages $V_{gc}$ and $V_{gs}$ to vary $T_p$ and $T_f$ by a  significant amount (see e.g. Fig.\ref{Fig-04}). For a $L \sim 1 \, {\rm \mu m}$ long junction, this range   is a few ${\rm meV}$. These values are well below the bulk gap and are consistent with the typical experimental conditions~\cite{TM-1-exp}, so that $T_p$ and $T_f$ should be  tunable in these regimes and display the internal structure of the junction.\\

{\it Applications in spintronics.} Our results   suggest  the possibility to exploit the setup in Fig.\ref{Fig-01} as a building block for spintronics nanodevices. The underlying idea is the following: due to  the factorization property shown here, for any given device in Fig.\ref{Fig-01} one can first determine $T_p$ and $T_f$ through transconductance measurements as described above. In particular, one can extract the dependence of $T_p$ and $T_f$  on the related gate voltage $V_{gc}$ and $V_{gs}$, respectively. Such `spectrum' of $T_p$ and $T_f$  depends on the specific features of the tunnel junction (and possibly on the presence of disorder) and represents a sort of fingerprints of the   tunnel junction. Then, by tuning the values of $V_{gc}$ and $V_{gs}$ according  to the obtained fingerprints, one can  operate electrically on $T_p$ and $T_f$,  independently, thereby realizing a device where spin-polarized currents can be steered and redistributed in the four terminals. 
\\

{\it Effects of electron-electron interaction.}  Although the analysis of the interacting case is beyond the purpose of the present paper, we would like to briefly discuss this aspect, addressing the main underlying  issues. Theoretical predictions show that, in the presence of electron-electron interaction, a pair of helical edge states  realize a helical Luttinger liquid (LL) where, besides a repulsively interacting charge sector characterized by a   Luttinger parameter $g<1$, the spin sector is also interacting with an attractive strength $1/g$~\cite{zhang-PRL,teo,chamon_2009,trauz-recher,strom_2009} ($g=1$ corresponds to the non interacting limit). This unconventional Luttinger liquid is thus particularly interesting and, despite  no clear experimental evidence of interaction effects on helical edge states has been observed so far, the search for conditions where these effects can be disguised is a fascinating problem. When a tunnel junction is realized, electron-electron interaction interplays with tunneling in a non-trivial way, leading to qualitatively different features as compared to the noninteracting case. In the limit of a short junction, and for $1/2 < g < 1$ (a range that includes the experimentally plausible regime $g \lesssim 1$ of weak interaction) the analysis based on the the DT model~\cite{teo,chamon_2009,strom_2009} shows that {p}- and {f}-tunneling terms are both irrelevant operators, with the same scaling dimension, due to the $g \leftrightarrow 1/g$ relation between charge and spin sector. Corrections to the ideal conductance are thus due to the finite bias voltage and/or  temperature, and appear as a power law  with a $g$-dependent exponent. However, when the finite length~$L$ of the junction is taken into account, the problem becomes intrinsically more complicated for various reasons. In the first instance, besides tunneling terms, also inter-edge forward interaction terms arise along the junction region, similarly to a spinful LL, breaking the $g \leftrightarrow 1/g$ relation that holds away from the junction~\cite{tanaka-nagaosa,trauz-recher}. As a consequence, {p}- and {f}-tunneling terms acquire different scaling dimensions and, even to lowest order in tunneling, qualitative modifications in the bias voltage dependence of the conductance are expected as compared to the interacting DT model. In the case of stronger tunneling $|\Gamma_\nu L|/\hbar v_F \sim 1$, these modifications may even be more significant because  of the `dressing' of each DT tunneling amplitude  by higher order contributions stemming from the other tunneling channel, similarly to  the non interacting case  Eq.(\ref{gammafDT-res}). Furthermore, inter-edge interaction also involves two types of 2-particle backward scattering, which preserve and flip spin, respectively~\cite{tanaka-nagaosa,teo,trauz-recher}. Finally, while along a non-interacting edge Rashba spin-orbit coupling cannot induce single-particle backscattering, in the presence of electron-electron interaction such coupling can lead to an effective 2-particle backscattering along the edge~\cite{crepin_2012,johannesson_2010}. In a tunnel junction, such intra-edge effect is expected to interplay with inter-edge tunneling and interaction. The whole problem can thus be formulated in terms of two coupled Luttinger liquids (for charge and spin sectors) with inhomogeneous interaction parameters $g_c(x)$ and $g_s(x)$, and with inhomogenous non-linear coupling arising from both tunneling and interaction terms, over the junction length. This highly non-trivial problem does not have an exact solution is general, and deserves a specific analysis. On the basis of  results obtained in some specific cases and on formal similarities with inhomogeneous LL in the presence of impurities~\cite{dolcetto-sassetti,dolcini_2003}, we can formulate some expectations for the case of weak tunneling $|\Gamma_\nu L|/\hbar v_F \ll 1$. In this case the conductances are modified with respect to the DT model by a modulation factor $f$, which depends in a non-monotonous way on the the ratio $eV/E_L$ between the bias voltage $V$ and  the energy scale $E_L \propto v_F/L$ associated to $L$. Also, the period of the oscillations shown e.g. in Fig.\ref{Fig-04} should be modified by an interaction-dependent factor.  Whether the factorization persists in the presence of interaction is a  challenging question.

%%%%%%%%%%%%%%%%%%%%%%%%%%%%%%%%%%%%%%%%%%%%%%%%%%%%%%%%%%%%%%%%%%%%%%%%%%%%%%%%%%%%
%%%%%%%%%%%%%%%%%%%%%%%%%%%%%%%%%%%%%%%%%%%%%%%%%%%%%%%%%%%%%%%%%%%%%%%%%%%%%%%%%%%%
%%%%%%%%%%%%%%%%%%%%%%%%%%%%%%%%%%%%%%%%%%%%%%%%%%%%%%%%%%%%%%%%%%%%%%%%%%%%%%%%%%%%
%%%%%%%%%%%%%%%%%%%%%%%%%%%%%%%%%%%%%%%%%%%%%%%%%%%%%%%%%%%%%%%%%%%%%%%%%%%%%%%%%%%%
%%%%%%%%%%%%%%%%%%%%%%%%%%%%%%%%%%%%%%%%%%%%%%%%%%%%%%%%%%%%%%%%%%%%%%%%%%%%%%%%%%%%
%%%%%%%%%%%%%%%%%%%%%%%%%%%%%%%%%%%%%%%%%%%%%%%%%%%%%%%%%%%%%%%%%%%%%%%%%%%%%%%%%%%%
%%%%%%%%%%%%%%%%%%%%%%%%%%%%%%%%%%%%%%%%%%%%%%%%%%%%%%%%%%%%%%%%%%%%%%%%%%%%%%%%%%%%
%%%%%%%%%%%%%%%%%%%%%%%%%%%%%%%%%%%%%%%%%%%%%%%%%%%%%%%%%%%%%%%%%%%%%%%%%%%%%%%%%%%%
%%%%%%%%%%%%%%%%%%%%%%%%%%%%%%%%%%%%%%%%%%%%%%%%%%%%%%%%%%%%%%%%%%%%%%%%%%%%%%%%%%%%
%%%%%%%%%%%%%%%%%%%%%%%%%%%%%%%%%%%%%%%%%%%%%%%%%%%%%%%%%%%%%%%%%%%%%%%%%%%%%%%%%%%%
%%%%%%%%%%%%%%%%%%%%%%%%%%%%%%%%%%%%%%%%%%%%%%%%%%%%%%%%%%%%%%%%%%%%%%%%%%%%%%%%%%%%
%%%%%%%%%%%%%%%%%%%%%%%%%%%%%%%%%%%%%%%%%%%%%%%%%%%%%%%%%%%%%%%%%%%%%%%%%%%%%%%%%%%%
\acknowledgments
The authors greatly acknowledge the German-Italian Vigoni Program for financial support, and B. Trauzettel and H. Buhmann for fruitful discussions. F.D. also  acknowledges financial support from FIRB 2012 project HybridNanoDev (Grant No.RBFR1236VV).

\end{document}